\documentclass[%
twocolumn,
superscriptaddress,
groupedaddress,
nofootinbib,
amsmath,amssymb,
aps,
pra,
]{revtex4}
\usepackage{amsmath}
\usepackage{empheq}
\usepackage{amssymb}
\usepackage{color}
\usepackage{graphicx}
\usepackage{algorithm}
\usepackage{algorithmicx}
\usepackage[noend]{algpseudocode}
\usepackage[caption=false]{subfig}
\usepackage{dcolumn}
\usepackage{bm}
\usepackage{braket}
\usepackage{fouridx}
\usepackage{times}
\usepackage[
colorlinks=true,
linkcolor=blue,
citecolor=blue,
urlcolor=blue,
]{hyperref}

\def\v#1{\mbox{\boldmath $#1$}}

\begin{document}

\preprint{APS/123-QED}
\title{Griffiths-McCoy singularity on the diluted Chimera graph:\\
Monte Carlo simulations and experiments on the quantum hardware}

\author{Kohji Nishimura}
\thanks{Present address: Jij Inc., Bunkyo, Tokyo 113-0031, Japan}
\email{knishimura@j-ij.com}
\affiliation{Department of Physics, Tokyo Institute of Technology, Yokohama, Kanagawa 226-8503, Japan}

\author{Hidetoshi Nishimori}
\affiliation{Institute of Innovative Research, Tokyo Institute of Technology, Yokohama, Kanagawa 226-8503, Japan}
\affiliation{Graduate School of Information Sciences, Tohoku University, Sendai, Miyagi 980-8579, Japan}
\affiliation{RIKEN Interdisciplinary Theoretical and Mathematical Sciences Program (iTHEMS), Wako, Saitama 351-0198, Japan}

\author{Helmut G. Katzgraber}
\affiliation{Microsoft Quantum, Microsoft, Redmond, Washington 98052, USA}

\date{\today}
\begin{abstract}
The Griffiths-McCoy singularity is a phenomenon characteristic of low-dimensional disordered quantum spin systems, in which the magnetic susceptibility shows singular behavior as a function of the external field even within the paramagnetic phase. We study whether this phenomenon is observed in the transverse-field Ising model with disordered ferromagnetic interactions on the quasi-two-dimensional diluted Chimera graph both by quantum Monte Carlo simulations and by extensive experiments on the D-Wave quantum annealer used as a quantum simulator.  From quantum Monte Carlo simulations, evidence is found for the existence of the Griffiths-McCoy singularity in the paramagnetic phase.  The experimental approach on the quantum hardware produces results that are less clear-cut due to the intrinsic noise and errors in the analog quantum device but can nonetheless be interpreted to be consistent with the existence of the Griffiths-McCoy singularity as in the Monte Carlo case. This is the first experimental approach based on an analog quantum simulator to study the subtle phenomenon of Griffiths-McCoy singularities in a disordered quantum spin system, through which we have clarified the capabilities and limitations of the D-Wave quantum annealer as a quantum simulator.
\end{abstract}

\maketitle

\section{Introduction}

Spin systems with disorder exhibit a number of unusual properties,  a most notable example of which is the spin-glass state with randomly frozen spin configurations caused by disorder and frustration \cite{nishimori01}.  Even in the absence of frustration, disorder alone can lead to unexpected behaviors, and the Griffiths singularity is one of the most prominent examples \cite{Vojta2010}.  It was shown by Griffiths \cite{Griffiths1969} that the magnetic susceptibility of a randomly diluted ferromagnetic system shows singularities as a function of the external magnetic field within the paramagnetic phase.  This singularity originates in the existence of rare but extremely large ferromagnetic clusters which behave almost like a pure ferromagnetic system responding strongly to the external field if the temperature is below the transition temperature of the non-diluted ferromagnetic system. The Griffiths singularity in the susceptibility is, however, a very weak essential singularity and is hard to observe experimentally.

Quantum effects significantly enhance the strength of this singularity, leading to divergences of the linear and nonlinear susceptibilities \cite{Vojta2010}, a phenomenon known as the Griffiths-McCoy singularity \cite{McCoy1968a,McCoy1968b}. Numerical, theoretical, and experimental investigations have been conducted on this subject, particularly in low-dimensional systems where the singular behavior of physical quantities is expected to appear most prominently \cite{RiegerYoung1994,RiegerYoung1996a,RiegerYoung1996b,Guo1994,Guo1996,Pich1998,Vojta2010,Singh2017,Wang2017}.

In the present paper we study this problem in the ferromagnetic transverse-field Ising model on the diluted quasi-two-dimensional Chimera graph with disordered interactions by quantum Monte Carlo simulation on a classical computer and by quantum hardware simulations on the D-Wave quantum annealer. There are several reasons to motivate this direction of investigation.  First, there exist few numerical or theoretical studies on the Griffiths-McCoy singularity for (quasi-)two-dimensional disordered ferromagnets without frustration \cite{Pich1998} although spin-glass systems have been investigated relatively extensively \cite{RiegerYoung1994,RiegerYoung1996a,RiegerYoung1996b,Guo1994,Guo1996,Singh2017}.  One expects that the disordered ferromagnet and spin glasses would behave qualitatively in the same way as far as the Griffiths-McCoy singularity is concerned because only disorder is relevant to this phenomenon, not the existence of frustration, but it nevertheless makes sense to confirm this conjecture explicitly numerically and experimentally by an additional concrete example.  Second, it is an interesting exercise to compare the results of numerical simulations with data from the analog quantum simulator, the latter of which can be regarded as an experimental apparatus because real physical phenomena corresponding to the theory are expected to take place on the chip of the device if the latter operates as designed.

We find that the Griffiths-McCoy singularity exists in the present problem from numerical simulations.  In contrast, data from the D-Wave device include a good amount of uncertainties due to noise, systematic bias, and other imperfections, but the results can be understood to be compatible with the existence of Griffiths-McCoy singularity, providing an experimental test of the existence of this subtle singularity on the analog quantum simulator.

This paper is organized as follows.
We summarize basic facts about the Griffiths-McCoy singularity and list two quantities to be measured, linear and nonlinear susceptibilities, by numerical and experimental approaches in Sec. \ref{sec:griffiths}. Methods and results of numerical simulations by quantum Monte Carlo are described in Sec. \ref{sec:qmc}. Experiments on the D-Wave device are discussed, and results are compared with those from numerical simulations in Sec. \ref{sec:dwave}.  Section \ref{sec:discussion} discusses the results.   Additional details are described in Appendixes.

\section{Griffiths-McCoy singularity}
\label{sec:griffiths}

In this section, we first explain the basic idea of the original Griffiths singularity for the classical ferromagnetic Ising model on a randomly diluted lattice, and then discuss the quantum version, the Griffiths-McCoy singularity for the transverse-field Ising model.

Let us consider the classical ferromagnetic Ising model on a regular lattice, e.g., the  two-dimensional square lattice, with a finite transition temperature between the paramagnetic and ferromagnetic phases. Suppose that we remove each bond (interaction) randomly with probability $1-p$ and keep the bond with probability $p$. The ferromagnetic phase becomes gradually unstable as $p$ decreases from 1 and the transition temperature $T_{c}(p)$ decreases as $p$ decreases. At the percolation threshold $p_{\rm c}$ for the given lattice, e.g., $p_{\rm c}=1/2$ for the square lattice, the transition temperature reaches zero, $T_{\rm c}(p_{\rm c})=0$.

Griffiths \cite{Griffiths1969} proved that, in the temperature range within the paramagnetic phase but below the transition temperature at $p=1$, $T_{\rm c}(p)<T<T_{\rm c}(1)$, the magnetic susceptibility $\chi (h)$ as a function of the external field $h$ is singular at $h=0$. This behavior exists for any $0<p<1$ both above and below the percolation threshold $p_{\rm c}$.  The intuitive reason is that there exist very large clusters of almost ferromagnetically-ordered spins even for $p<p_{\rm c}$ and they respond strongly to the external field if $T<T_{\rm c}(1)$ because, in the infinitely large system,  the spontaneous magnetization is positive for $h>0$ and is negative for $h<0$, i.e., a discontinuity at $h=0$ (an infinite susceptibility).  Since the probability of the existence of very large clusters is exponentially small as a function of their size, the resulting Griffiths singularity in the susceptibility is very weak, an essential singularity, and is therefore hard to detect experimentally. 

Introduction of quantum effects significantly enhances the strength of the singularity, known as the Griffiths-McCoy singularity \cite{McCoy1968a,McCoy1968b,Vojta2010}. 
Let us formulate the statistical mechanics of the transverse-field Ising model
\begin{align}
    H = \sum_{\braket{i,j}} J_{ij}\hat{\sigma}_i^z \hat{\sigma}_j^z - \Gamma \sum_{i}\hat{\sigma}_i^x
    \label{eq:TFIM}
\end{align}
in terms of the Suzuki-Trotter decomposition \cite{Suzuki1976}.
This is a standard method for classical simulation (quantum Monte Carlo) of the transverse-field Ising model, in which the quantum problem is mapped to a collection of classical paths, as in the Feynman path integral, along discretized imaginary time steps. In this classical representation of the transverse-field Ising model,
Ising spins along the Trotter (imaginary time) axis are coupled strongly by ferromagnetic interactions for moderate and weak values of the transverse field $\Gamma$.  This strong ferromagnetic coupling along the Trotter direction causes the ferromagnetically coupled clusters extended in the Trotter direction, greatly enhancing the response to the external field. Indeed, at zero temperature, the linear and nonlinear susceptibilities $\chi$ and $\chi_{\mathrm{nl}}$ are known to behave as \cite{Vojta2010}
\begin{align}
    \label{eq:griffiths:chi1}
    \chi &\simeq h^{d/z'-1}, \\
    \label{eq:griffiths:chi2}
    \chi_{\mathrm{nl}} &\simeq h^{d/z'-3}, 
\end{align}
where $d$ is the spatial dimension of the lattice and $z'$ is effective (running) dynamical exponent dependent on the value of $\Gamma$.  Equations (\ref{eq:griffiths:chi1}) and (\ref{eq:griffiths:chi2}) indicate that $\chi$ and $\chi_{\mathrm{nl}}$ diverge at $h=0$ if $d/z'<1$ and $d/z'<3$, respectively.  The former inequality is indeed satisfied on the two-dimensional square lattice with the interaction and the transverse field chosen uniformly randomly, and the exponent $z'$ diverges at the transition point \cite{Pich1998}. In one dimension, the singularity is stronger with $z'$ being divergent for any $\Gamma$ larger than the transition point \cite{Fisher1995}. In three dimensions, if the singularity exists, it is weak \cite{Singh2017}. These properties hold for other types of disorder in the interactions, not just simple dilution of ferromagnetic interactions, including the Ising spin glass model under a transverse field, because only the existence of randomness is relevant in the sense of renormalization group and frustration does not play an essential role in the Griffiths-McCoy singularity.

In order to measure $d/z'$ to determine whether or not the linear and nonlinear susceptibilities diverge, it is useful to plot the histogram of measured values of local linear and nonlinear susceptibilities, which are known to behave as  \cite{RiegerYoung1996a,RiegerYoung1996b,Guo1996}
\begin{align}
    \label{eq:griffiths:logplocchi}
    \ln P(\chi_{\mathrm{loc}}) \simeq -\left(\frac{d}{z'}+1\right) \ln \chi_{\mathrm{loc}}, \\
    \label{eq:griffiths:logpnllocchi}
    \ln P(\chi_{\mathrm{nlloc}}) \simeq -\left(\frac{d}{3z'}+1\right) \ln \chi_{\mathrm{nlloc}},
\end{align}
which holds for large values of the variables.  The slope of the plot gives the exponent. The local susceptibility $\chi_{\mathrm{loc}}$ is defined as
\begin{align}
    \chi_{\mathrm{loc}}=\chi_{ii}= \left.\frac{\partial m_i}{\partial h_i}\right|_{h_i=0},
\end{align}
where $m_i$ is the local magnetization computed from the quantum-mechanical expectation value of the spins and $h_i$ is the local longitudinal field applied only to site $i$. The local susceptibility $\chi_{ii}$ depends on the site index $i$, and we collect the statistics of this quantity over $i$ and random samples to generate the histogram $P(\chi_{\mathrm{loc}})$. The local nonlinear susceptibility $\chi_{\mathrm{nlloc}}$ is defined similarly as the third derivative of $m_i$ with respect to $h_i$.  In the paramagnetic phase, the correlation length is short and the ordinary (global) linear and nonlinear susceptibilities are expected to follow the same formulas as Eqs.~(\ref{eq:griffiths:logplocchi}) and (\ref{eq:griffiths:logpnllocchi}).  The local quantities are used often in numerical simulations because a larger number of samples can be generated than for the global quantities, leading to better statistics.

\section{Quantum Monte Carlo simulation}
\label{sec:qmc}
\begin{figure}[tbp]
	\centering
    \includegraphics[width=.6\linewidth]{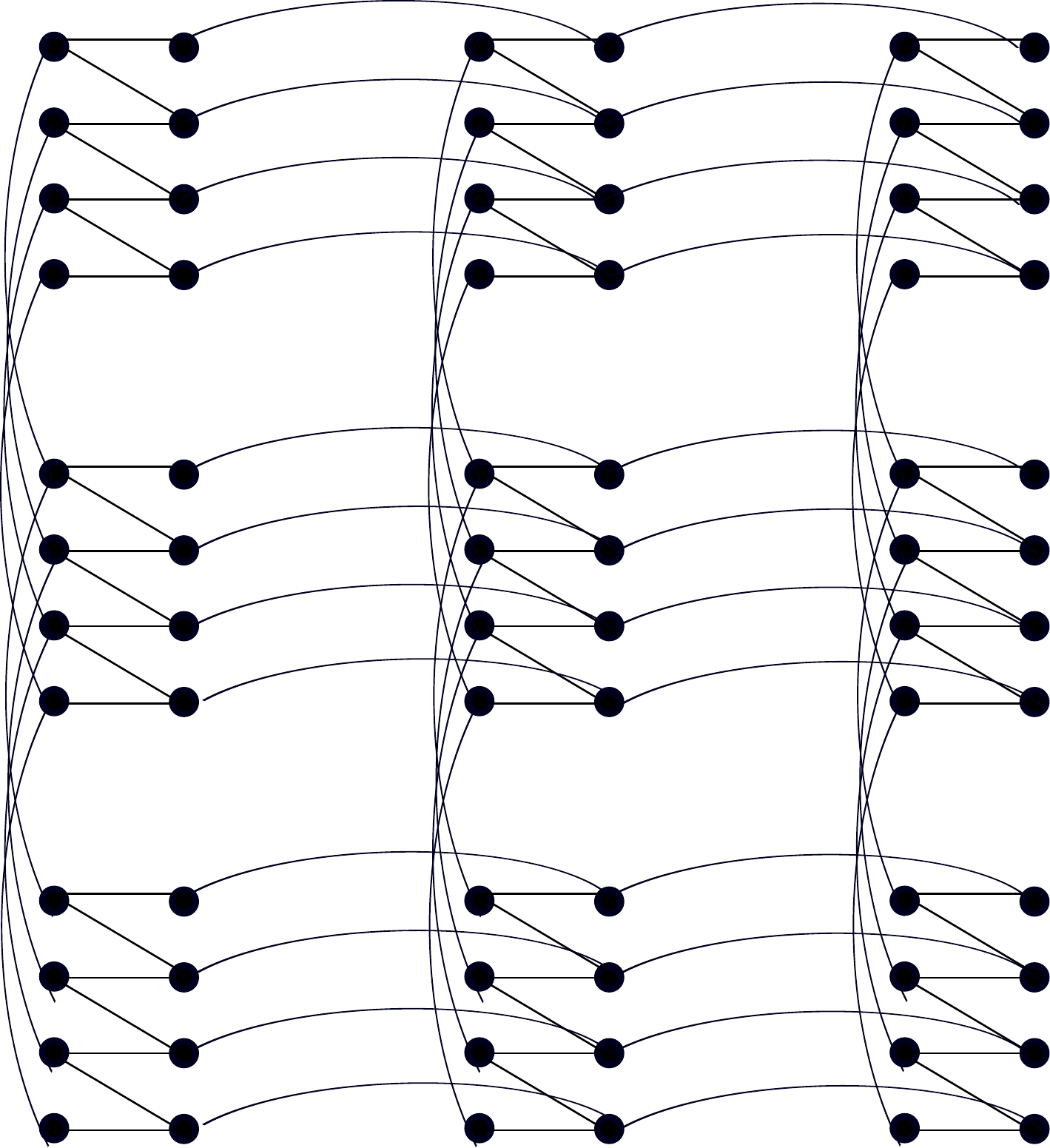}
    \caption{Connectivity in the diluted Chimera graph. The original Chimera graph has much more interactions. For example, the site on the upper-left corner has interactions with all four sites within the unit cell of 4+4 sites but only two of them remain here.}
	\label{fig:qmc:sparseChimera}
\end{figure}

In this section we describe the methods and results of quantum Monte Carlo simulations of the Griffiths-McCoy singularity on the diluted Chimera graph as depicted in Fig.~\ref{fig:qmc:sparseChimera}, in which some of the interactions on the original Chimera graph are removed.
This dilution of the Chimera graph is expected to help us enhance the parameter range of the Griffiths-McCoy singularity since lower-dimensional systems with less degree of connectivity tend to show stronger effects of this singularity \cite{Singh2017}.  The Chimera graph is chosen because the topology is directly realized on the D-Wave chip without the necessity of embedding.  The above dilution of the Chimera graph is not random, and it has no effect on the Griffiths-McCoy singularity except for the enhancement of parameter range as written above.

Randomness causing the singularity is in the choice of the interaction strength at the bonds remaining on the graph of Fig.~\ref{fig:qmc:sparseChimera},
\begin{align}
    P(J_{ij}) = \frac{1}{6}\sum_{k=0}^{5}\delta(J_{ij} + 0.2k),
    \label{eq:Jij_QMC}
\end{align}
meaning that the $J_{ij}$ are uniformly chosen from the set $\{0, -0.2, -0.4, -0.6, -0.8, -1.0\}$. This problem on the quasi-two-dimensional diluted Chimera graph has not been studied so far, and it is interesting to investigate it both by classical simulation, i.e., quantum Monte Carlo, and by a direct quantum simulation on the D-Wave quantum annealer, which may be regarded as an experiment on the analog quantum simulator. The present section concerns the former classical simulation.

\subsection{Method}

We perform quantum Monte Carlo simulations based on the Suzuki-Trotter decomposition with parameters listed in Table \ref{tab:qmc:qmcparam}.
\begin{table*}[ht]
    \centering
    \caption{Parameters for the quantum Monte Carlo simulation. $N_\mathrm{rand}$ is the number of random instances, $M$ is the number of Trotter slices, $L$ is the square root of the number of Chimera units ($N = 8L^2$ is the total number of sites), $N_\mathrm{step}$ is the number of Monte Carlo steps, $\beta_{\mathrm{max}}$ and $\beta_{\mathrm{min}}$ are, respectively, the maximum and minimum values of the inverse temperature, $N_\beta$ is the number of inverse temperatures, $\Gamma_{\mathrm{min}}$  and $\Gamma_{\mathrm{max}}$ are, respectively, the minimum and maximum values of the transverse field, and $N_\Gamma$ is the number of transverse field values.}
    \label{tab:qmc:qmcparam}
    \begin{tabular*}
    {\columnwidth}{@{\extracolsep{\fill}} c c c c c c c c c c}
\hline
\hline
    $N_\mathrm{rand}$ & $M$ & $L$ &$N_\mathrm{step}$ &$\beta_{\mathrm{max}}$ & $\beta_{\mathrm{min}}$ & $N_\beta$ &$\Gamma_{\mathrm{min}}$ &$\Gamma_{\mathrm{max}}$ & 
    $N_\Gamma$\\
    \hline
    200& 150 & 6,8,10,12 & $2^{20}$& 50 & 2.5 & 10 & 1.4 & 2.135& 50\\
  \hline
\end{tabular*}
\end{table*}
The total number of spins we deal with for each Monte Carlo step is $L\times L \times 8 \times M \times N_{\beta} \times N_{\Gamma}$. See Table \ref{tab:qmc:qmcparam} for the definition of each symbol. For each random instance, we store the following physical variables: the absolute total magnetization, the squared total magnetization, the fourth moment of the total magnetization, the squared local magnetization, and the  fourth moment of the local magnetization,
\begin{itemize}
    \item  $\braket{|m|} = \Braket{\left|\frac{1}{NM}\sum_{i=1}^{N}\sum_{t=1}^{M}\sigma_i(t)\right|}$,
    \item  $\braket{m^2} = \Braket{\left(\frac{1}{NM}\sum_{i=1}^{N}\sum_{t=1}^{M}\sigma_i(t)\right)^2}$,
    \item $\braket{m^4} = \Braket{\left(\frac{1}{NM}\sum_{i=1}^{N}\sum_{t=1}^{M}\sigma_i(t)\right)^4}$,
    \item  $\braket{m_i^2} = \Braket{\left(\frac{1}{M}\sum_{t=1}^{M}\sigma_i(t)\right)^2}$,
    \item$\braket{m_i^4} = \Braket{\left(\frac{1}{M}\sum_{t=1}^{M}\sigma_i(t)\right)^4}$,
\end{itemize}
where the brackets $\braket{...}$ denote the statistical-mechanical average (i.e., the Monte Carlo average), which is expected to reduce to the quantum-mechanical average in the low-temperature limit, and $m$ stands for the sum of spin values over all spatial sites and along the Trotter direction,
\begin{align}
    m = \frac{1}{NM}\sum_{i=1}^{N}\sum_{t=1}^{M}\sigma_i(t).
\end{align}
The quantity $m_i$ is the spin of local site averaged over the Trotter direction,
\begin{align}
    m_i = \frac{1}{M}\sum_{t=1}^{M}\sigma_i(t).
\end{align}
We employ a GPU-based algorithm to accelerate the Monte Carlo simulations.

We apply finite-size scaling to the analysis of the critical point and critical exponents through the Binder ratio $g$ \cite{Landau_Binder}:
\begin{align}
    g = \left[\frac{1}{2}\left(3-\frac{\braket{m}^4}{\braket{m^2}^2}\right)\right],
\end{align}
the global susceptibility $\chi$,
\begin{align}
    \chi = \beta N \left[\braket{m^2}\right],
\end{align}
and the magnetization $\left[\braket{|m|}\right]$, where $\left[\cdots\right]$ denotes the average over instances of randomness in interactions.

As for the exponent $d/z'$, we generate the histograms $P(\chi_{\mathrm{loc}})$ ad $P(\chi_{\mathrm{nlloc}})$ of the local susceptibility $\chi_{\mathrm{loc}}$ and the nonlinear local susceptibility $\chi_{\mathrm{nlloc}}$,
\begin{align}
    \chi_{\mathrm{loc}} &= \braket{m_i^2}, \\
    \chi_{\mathrm{nlloc}} &= -\left(\braket{m_i^4} - 3\braket{m_i^2}^2\right).
\end{align}
These quantities are expected to show the behavior described in Eqs.~(\ref {eq:griffiths:logplocchi}) and (\ref{eq:griffiths:logpnllocchi}). Each histogram is generated from $N \times N_\mathrm{rand}$ samples, where $N=8L^2$ is the number of sites.

We also plot similar histograms for the global linear and nonlinear susceptibilities,
\begin{align}
    \chi &= \braket{m^2}, \\
    \chi_{\mathrm{nl}} &= -(\braket{m^4} - 3\braket{m^2}^2),
\end{align}
to compare their behavior with the corresponding data for local susceptibilities. Each histogram is generated from $N_\mathrm{rand}$ samples.
We use the Python module ``scipy.optimize.curve\_fit" in SciPy \cite{Scipy} for the data fittings.

\subsection{Results}
\label{sec:QMC_result}
We first determine the transition point between the paramagnetic and ferromagnetic phases and then move on to the Griffiths-McCoy singularity within the paramagnetic phase.

\subsubsection{Transition point}
\begin{figure}[tbp]
	\centering
    \includegraphics[width=.9\linewidth]{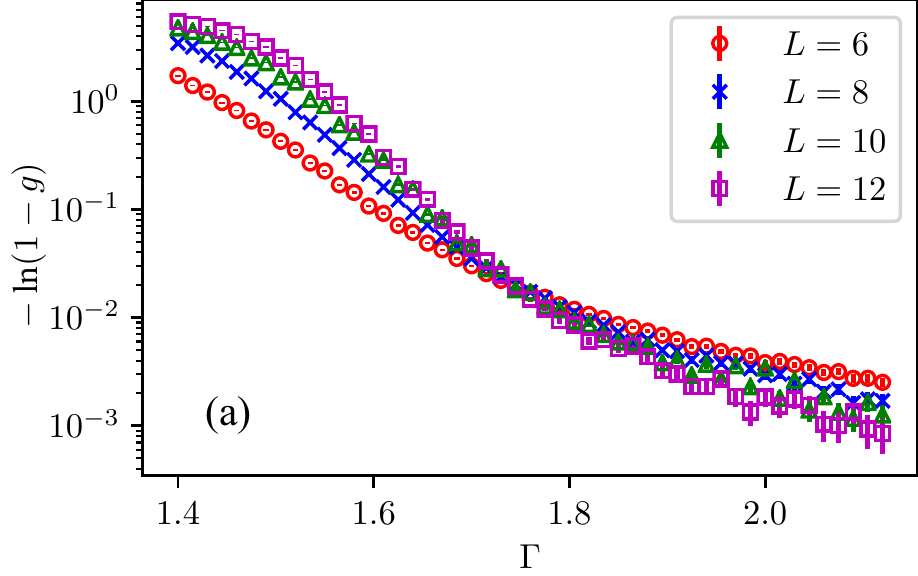}
    \includegraphics[width=.9\linewidth]{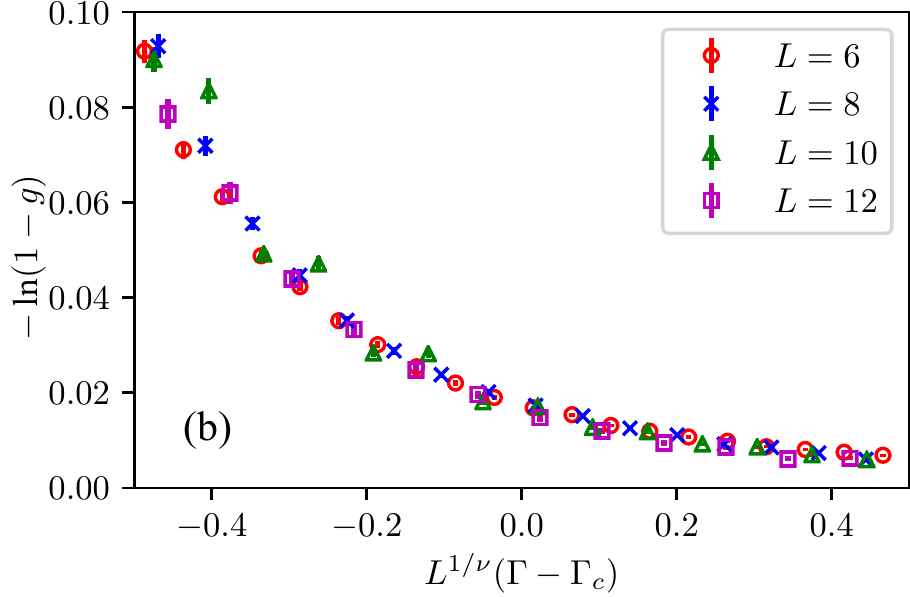}
    \caption{ (a) The Binder ratio $g$ in the scale $-\ln(1-g)$ for different system sizes $L=6, 8, 10, 12$ with the inverse temperature $\beta=20$. Curves cross at $\Gamma \simeq 1.7$. (b) Finite-size scaling analysis of the Binder ratio with $\Gamma_{\rm c} = 1.75(4)$ and $\nu = 1.4(2)$.}
	\label{fig:qmc:Binder}
\end{figure}

Figure \ref{fig:qmc:Binder}(a) shows the Binder ratio as a function of $\Gamma$ with the inverse temperature $\beta = 20$.
We employ $-\ln(1-g)$ instead of the naive Binder ratio $g$ in order to obtain a good resolution near the critical point \cite{LogBinder}. The figure indicates that the phase transition point is located around $\Gamma \simeq 1.7$. 

Figure \ref{fig:qmc:Binder}(b) shows the result of a finite-size scaling analysis of the same data. It is observed that the data with different sizes collapse onto the same curve for $\Gamma_{\rm c} = 1.75(4)$ and $\nu = 1.4(2)$.

\begin{figure}[tbp]
	\centering
    \includegraphics[width=.9\linewidth]{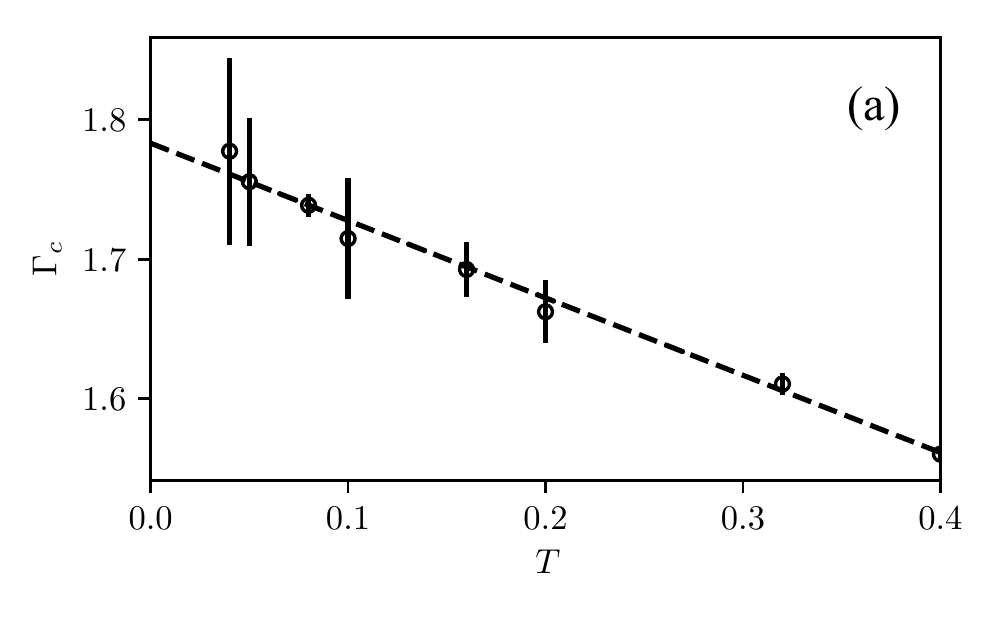}
    \includegraphics[width=.9\linewidth]{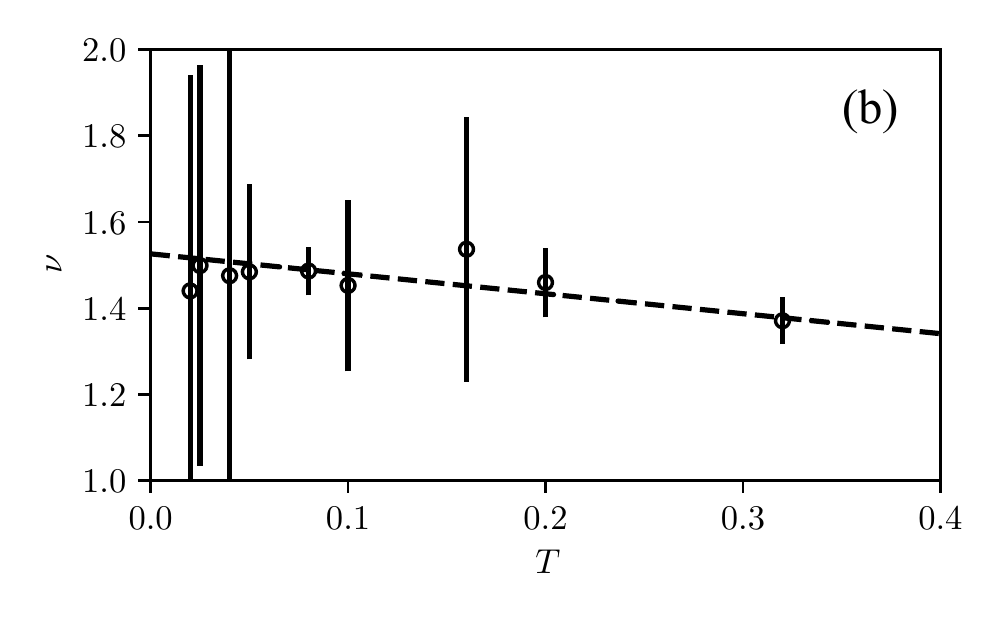}
    \caption{(a) Critical point $\Gamma_{\rm c}$ and (b) the critical exponent  $\nu$  as functions of the temperature $T$. Black dashed lines represent the linear fitting of the data. Extrapolation shows that $\Gamma_{\rm c} \simeq 1.78$ and $\nu \simeq 1.5$ in the zero temperature limit.}
	\label{fig:qmc:extrapolate}
\end{figure}

We apply the same finite-size scaling to each temperature, and the results for the transition point and the critical exponent are summarized in Fig.~\ref{fig:qmc:extrapolate}.
From Fig.~\ref{fig:qmc:extrapolate}(a), we observe that $\Gamma_{\rm c}$ grows linearly as the temperature $T$ approaches zero. Linear fitting shows that $\Gamma_{\rm c}$ can be determined to be about $1.78$ in the limit of zero temperature.
We also see that the critical exponent $\nu$ does not clearly depend on the temperature and can be estimated as $\nu \simeq 1.5$.
One may notice that the error bars are relatively large near zero temperature. This large uncertainty comes from the fact that the data collapse in Fig.~\ref{fig:qmc:Binder}(b) does not depend very much on the values of $\Gamma_{\rm c}$ and $\nu$: the data collapse is robust against the change of the values of $\Gamma_{\rm c}$ and $\nu$, especially the latter.  We think it reasonable to assume $\Gamma_{\rm c}=1.78(2)$ based on the data in the temperature range $T>0.1$ where the data are relatively stable. Estimation of $\nu$ and other critical exponents such as $\beta$ and $\gamma$ involves large uncertainties as was the case in three-dimensional spin glasses \cite{Katzgraber2006}, as detailed in Appendix \ref{sec:appendix_exponents}. At least, the estimated transition point $\Gamma_{\rm c}=1.78$ gives consistent results in finite-size scaling of other physical quantities such as the susceptibility and magnetization. See Appendix \ref{sec:appendix_exponents}.

\subsubsection{Histograms of susceptibilities}
\label{sec:qmc:hist}

In this section we estimate the exponent $d/z'$, which is critical to determine the existence of the Griffiths-McCoy singularity, from the data of local and global susceptibilities.
\subsubsection*{Local susceptibility}
\begin{figure}[tbp]
	\centering
    \includegraphics[width=.86\linewidth]{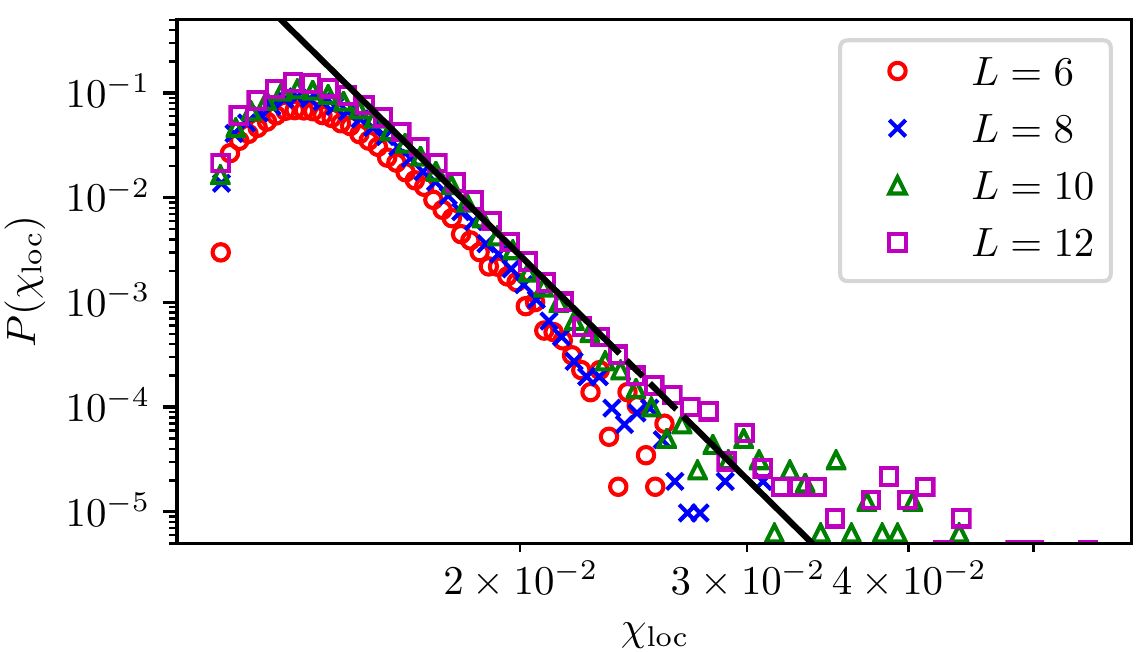}
    \caption{ Log-log plot of the histogram of the local susceptibility $P(\chi_\mathrm{loc})$ for different sizes at inverse temperature $\beta=50$ and transverse field $\Gamma = 1.895$. The black solid line shows a fit to the data whose slope is  $-13.83\pm 0.15$, which indicates that $d/z'\simeq 12.83\pm 0.15$.}
	\label{fig:qmc:lochistg1895}
\end{figure}

Figure \ref{fig:qmc:lochistg1895} shows the histogram of the local susceptibility $P(\chi_\mathrm{loc})$ in a paramagnetic region far from the critical point ($\Gamma = 1.895$).
In this region, the size dependence of the data is weak except for the tail of the distribution where the probability is small due to insufficient statistics. Since Eq.~(\ref{eq:griffiths:logplocchi}) is valid for large $\chi_{\mathrm{loc}}$, we have to carefully choose the range to fit the data to Eq.~(\ref{eq:griffiths:logplocchi}). Using the data in the range between slightly below the peak and $P(\chi_{\mathrm{loc}})\simeq 10^{-4}$,
$d/z'$ is estimated to be around $12.83$. 
According to the discussion in Sec. \ref{sec:griffiths}, we  conclude that both linear and nonlinear susceptibilities exhibit no divergence at this value $\Gamma=1.895$ since $d/z' > 3$.

\begin{figure}[tbp]
	\centering
    \includegraphics[width=.9\linewidth]{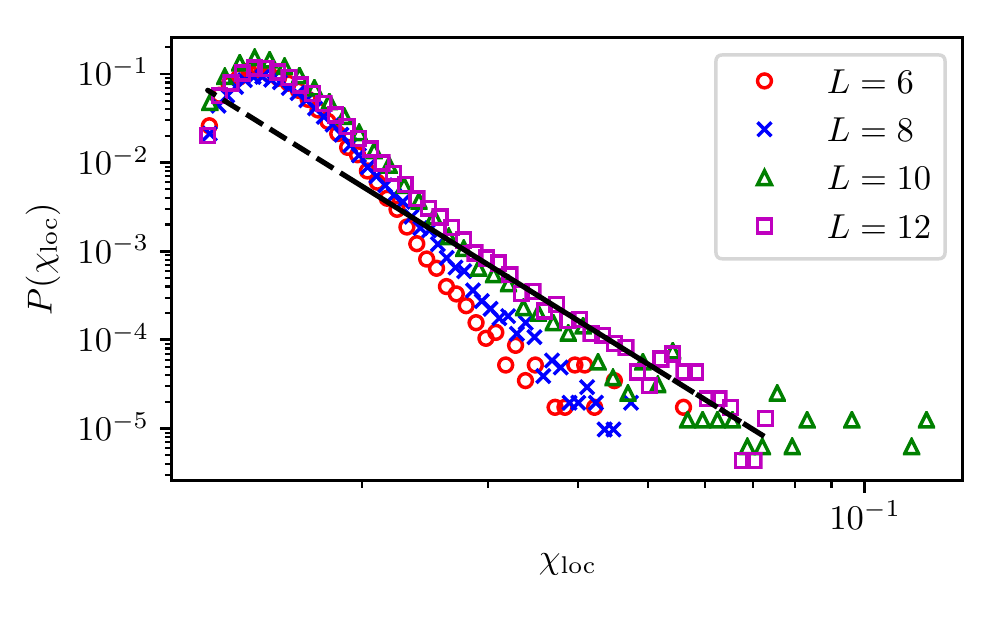}
    \caption{ Log-log plot of the histogram of the local susceptibility $P(\chi_\mathrm{loc})$ for different sizes at inverse temperature $\beta=50$ and transverse field $\Gamma = 1.79$. The slope clearly depends on the system size compared to Fig.~\ref{fig:qmc:lochistg1895}. The black solid line shows a fit to the data with the largest system size $L=12$ whose slope is $-5.40 \pm 0.33$, which indicates that $d/z'$ is at least smaller than about $4.40 \pm 0.33$.}
	\label{fig:qmc:lochistg179}
\end{figure}

Figure \ref{fig:qmc:lochistg179} is for $\Gamma=1.79$, closer to the critical point $\Gamma_{\rm c}=1.78$.
We observe that the slope clearly depends on the system size due to the finite-size effect and the slope tends to be shallower as the system size increases. To extract the information on $d/z'$, we use the data for the largest size $L=12$ since this is expected to give a lower bound of the slope in the large-size limit. We thus conclude that the exponent $d/z'$ is smaller than about $4.4$.

\begin{figure}[tbp]
	\centering
    \includegraphics[width=.9\linewidth]{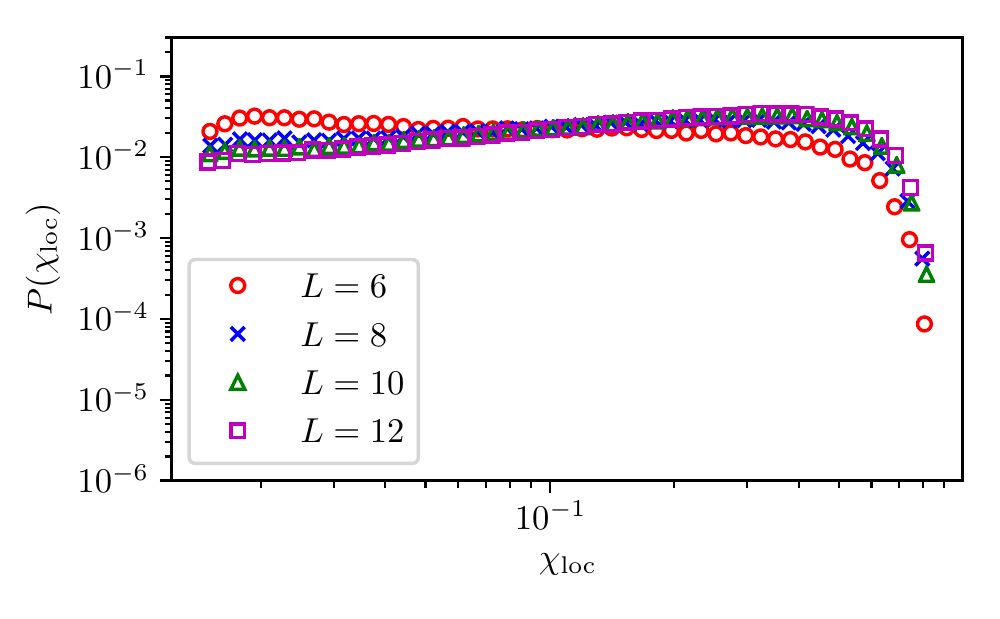}
    \caption{ Log-log plot of the histogram of the local susceptibility $P(\chi_\mathrm{loc})$ for different sizes at inverse temperature $\beta=50$ and transverse field $\Gamma = 1.4$ (ferromagnetic phase).  $P(\chi_{\mathrm{loc}})$ grows monotonically as the susceptibility $\chi_{\mathrm{loc}}$ increases especially for the largest system size $L = 12$.}
	\label{fig:qmc:lochistg14}
\end{figure}

The histogram $P(\chi_\mathrm{loc})$ for data in the ferromagnetic phase $\Gamma_{\rm c} = 1.4$ is in Fig.~\ref{fig:qmc:lochistg14}.
We find the behavior is quite different from previous cases for the paramagnetic phase. In the ferromagnetic phase, $P(\chi_{\mathrm{loc}})$ a slightly increasing function especially for the large system size.

\begin{figure}[tbp]
	\centering
    \includegraphics[width=.9\linewidth]{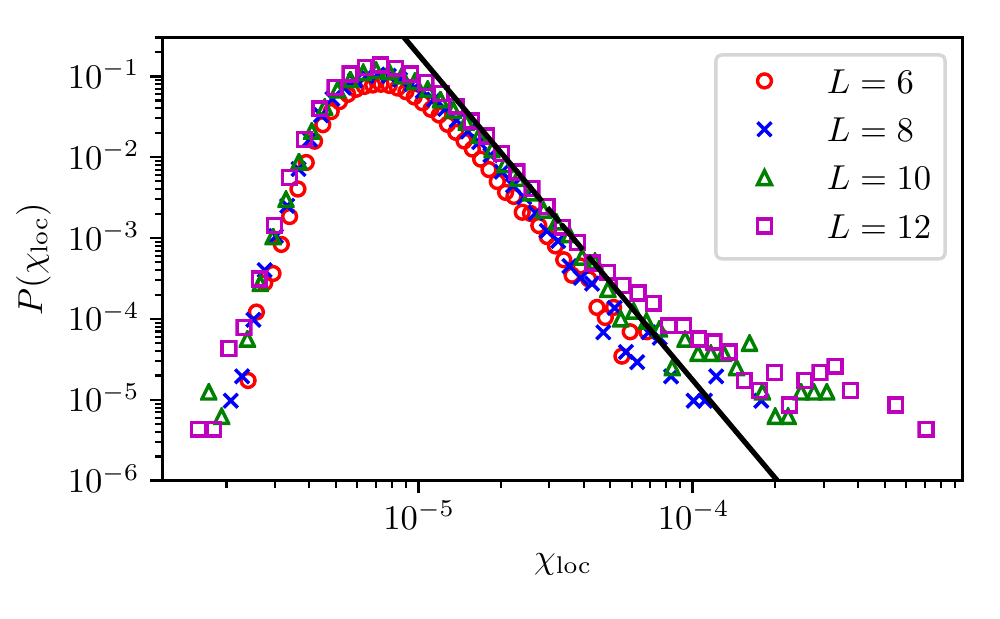}
    \caption{ Log-log plot of the histogram of the nonlinear local susceptibility $P(\chi_\mathrm{nlloc})$ for different sizes at inverse temperature $\beta=50$ and transverse field $\Gamma = 1.895$. The black solid line shows the fitting to the data whose slope is  $-4.02 \pm 0.06$, which indicates that $d/3z'$ is around $3.02 \pm 0.06$.}
	\label{fig:qmc:nlochistg1895}
\end{figure}

\begin{figure}[tbp]
	\centering
    \includegraphics[width=.9\linewidth]{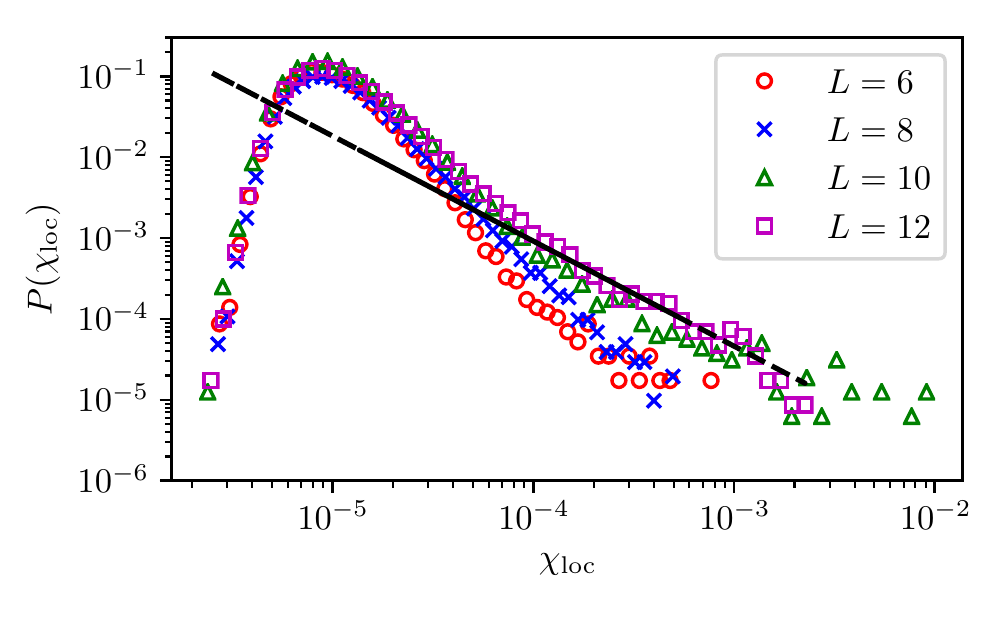}
    \caption{ Log-log plot of the histogram of the nonlinear local susceptibility $P(\chi_\mathrm{nlloc})$ for different sizes at inverse temperature $\beta=50$ and transverse field $\Gamma = 1.79$. The black solid line shows the fitting to the data with the largest system size $L=12$ whose slope is $-1.29 \pm 0.08$, which indicates that $d/3z'$ is smaller than $0.29 \pm 0.08$.}
	\label{fig:qmc:nlochistg179}
\end{figure}

We analyze the nonlinear local susceptibility $\chi_{\mathrm{nlloc}}$ similarly. The results for $\Gamma=1.895$ (far from the critical point) and $\Gamma = 1.79$ (near the critical point) are shown in Figs. \ref{fig:qmc:nlochistg1895} and \ref{fig:qmc:nlochistg179}, respectively.
The histograms exhibit similar behavior to that of the linear local susceptibility, where the plot has no size-dependence for the transverse field $\Gamma$ far from the critical point $\Gamma_{\rm c}$ and has clear size dependence for $\Gamma$ near $\Gamma_{\rm c}$. 

\begin{figure}[tbp]
	\centering
    \includegraphics[width=.9\linewidth]{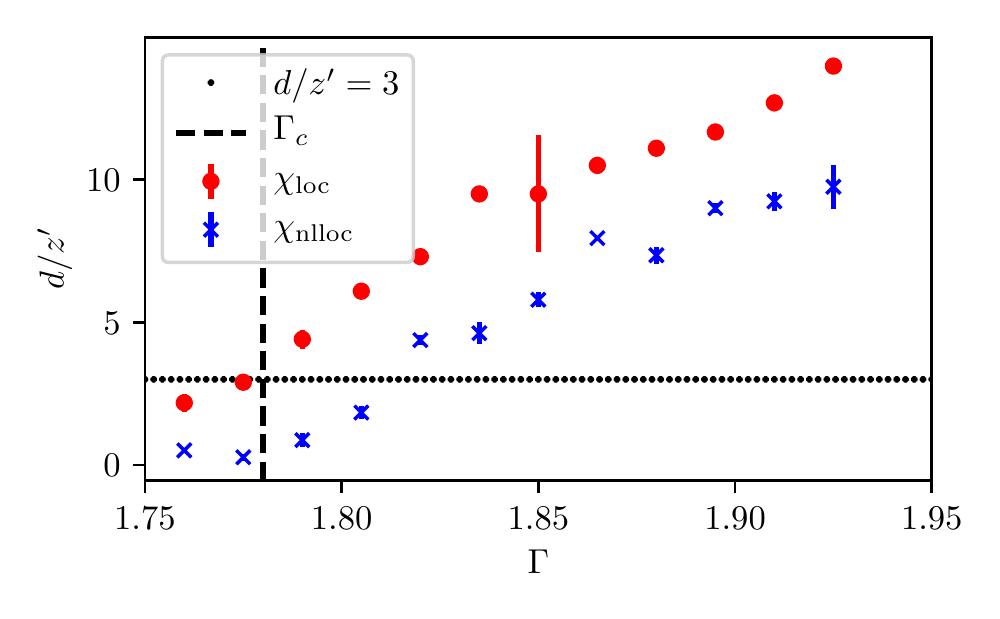}
    \caption{ 
    The exponent $d/z'$ as a function of $\Gamma$ measured at the inverse temperature $\beta = 50$. The data in circles and crosses are taken from $\chi_{\mathrm{loc}}$ and $\chi_{\mathrm{nlloc}}$, respectively. The vertical dashed line shows the critical point in the zero temperature limit $\Gamma_{\rm c} = 1.78$. The critical point corresponding to this finite-temperature ($\beta=50$) data would be smaller than $\Gamma_{\rm c}=1.78$ for the zero-temperature value.
    The horizontal dotted line is for $d/z'=3$, where the nonlinear susceptibility starts to diverge. Notice that the present data for $d/z'$ from finite-size simulations give upper bounds.}
	\label{fig:qmc:dzgamma}
\end{figure}

Extracted values of the exponent are plotted in Fig.~\ref{fig:qmc:dzgamma}.
Although there exist uncertainties in these values, it is useful to take into account the fact that the plotted values of the exponent $d/z'$ give upper bounds. We then observe the plausibility that $d/z'$ becomes smaller than the threshold $d/z'=3$ before the critical point is reached, suggesting that the nonlinear susceptibility diverges within the paramagnetic phase. This behavior is consistent with the previous study for the case of continuous distributions of random ferromagnetic interactions and random transverse field for a system on the square lattice \cite{Pich1998}. More subtle is the divergence of the linear susceptibility since it is difficult to determine from the data whether or not $d/z'$ becomes smaller than 1 in the paramagnetic phase $\Gamma>\Gamma_{\rm c}$.

\subsubsection*{Global susceptibility}
\begin{figure}[tbp]
	\centering
    \includegraphics[width=.9\linewidth]{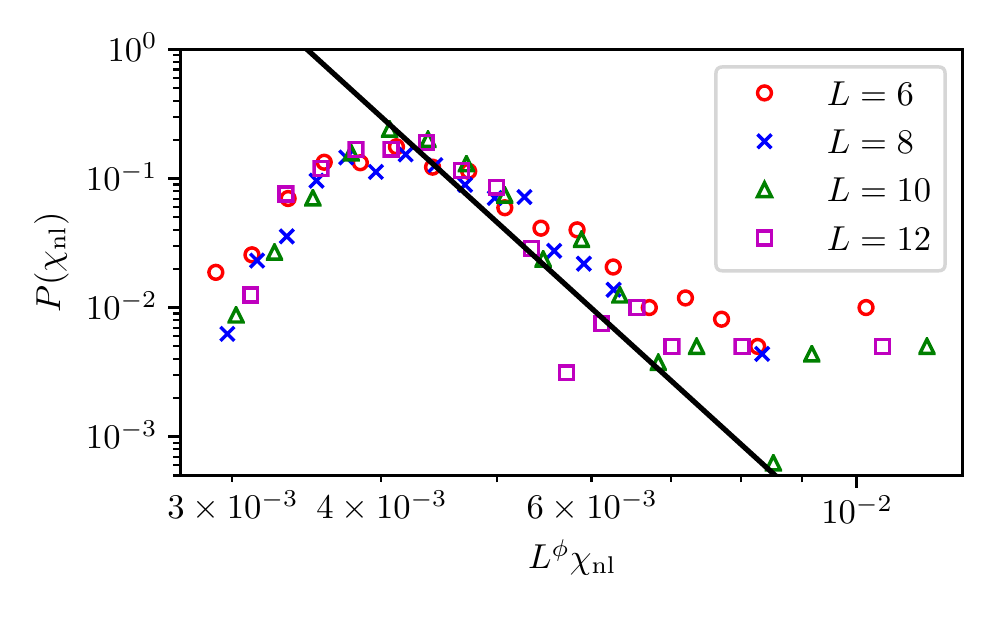}
    \caption{ Histogram of the global susceptibility at $\Gamma = 1.925$ (far from the critical point). The black solid line shows the fitting to the data with slope  $-8.42 \pm 1.8 $, indicating that $d/z'$ is about $7.42 \pm 1.8$.}
	\label{fig:qmc:glhistg1925}
\end{figure}

\begin{figure}[tbp]
	\centering
    \includegraphics[width=.9\linewidth]{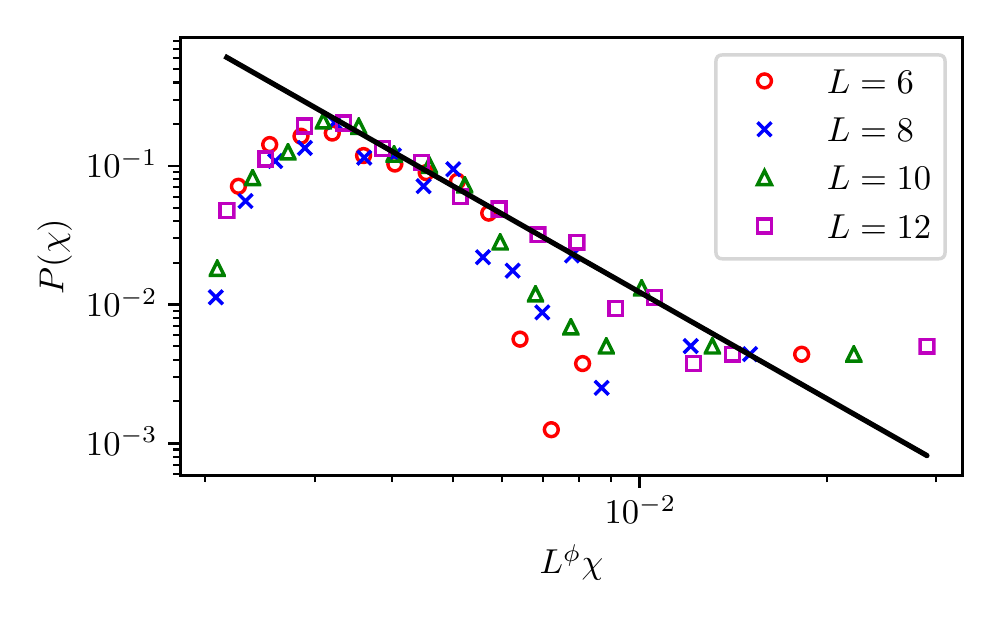}
    \caption{ Histogram of the global susceptibility at $\Gamma = 1.79$ (close to the critical point). The black solid line is the fitting to the data with slope $-2.5 \pm 0.22$, or $d/z'$ being around $1.5 \pm 0.22$.}
	\label{fig:qmc:glhistg179}
\end{figure}

\begin{figure}[tbp]
	\centering
    \includegraphics[width=.9\linewidth]{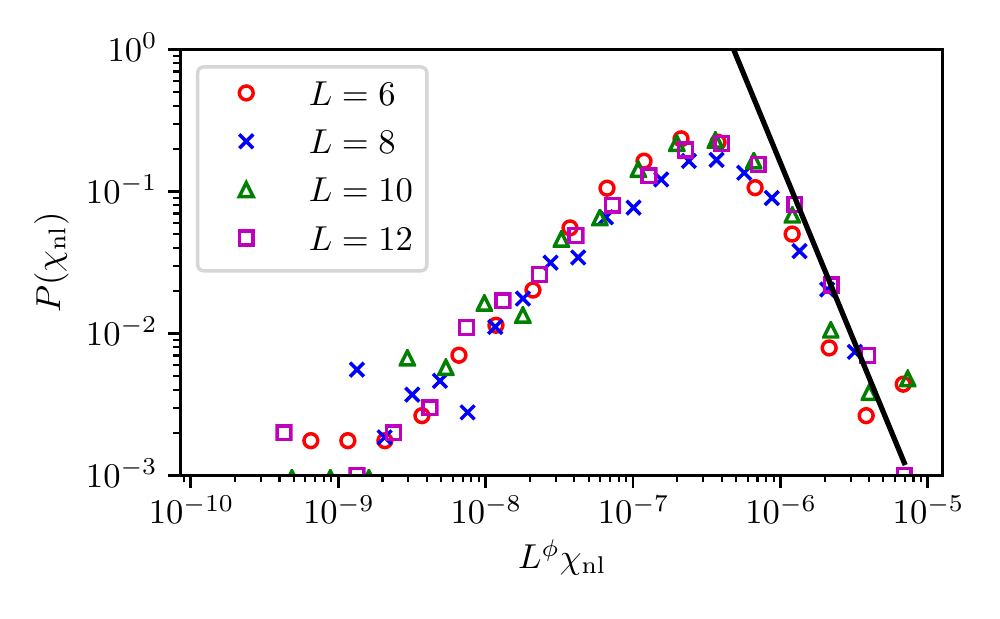}
    \caption{ Histogram of the global nonlinear susceptibility at $\Gamma = 1.91$. The black solid line is the fitting to the data with slope  $-2.5 \pm 0.25$, meaning that $d/3z'$ is around $1.5 \pm 0.25$.}
	\label{fig:qmc:nglhistg191}
\end{figure}

\begin{figure}[tbp]
	\centering
    \includegraphics[width=.9\linewidth]{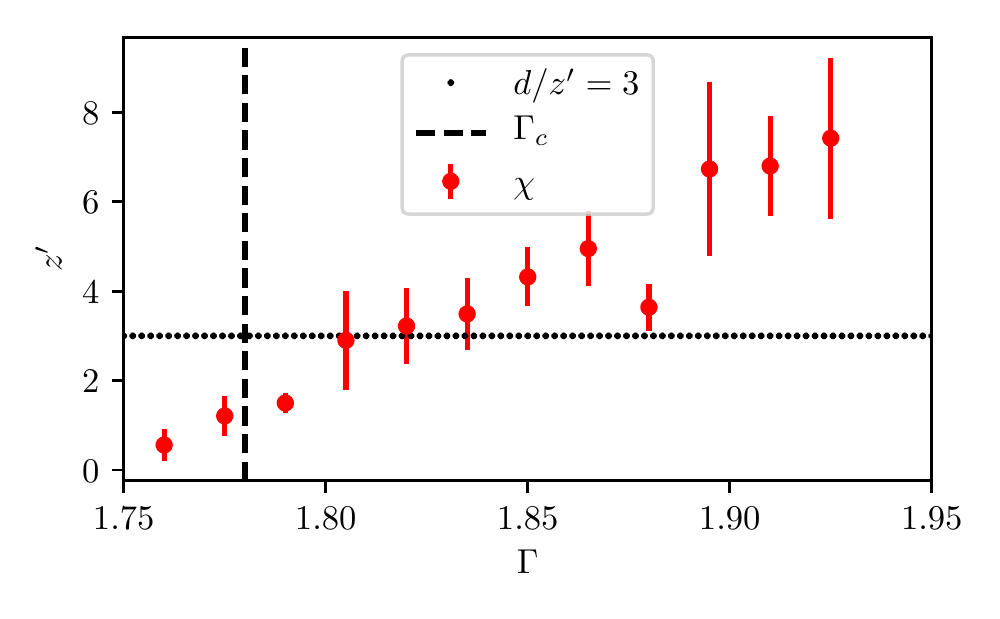}
    \caption{ Similar plot to Fig.~\ref{fig:qmc:dzgamma} but for the exponent extracted from the global linear susceptibility.}
	\label{fig:qmc:gldzgamma}
\end{figure}

We next verify if the data for the global susceptibility are consistent with those for the local susceptibility. Figures \ref{fig:qmc:glhistg1925}, \ref{fig:qmc:glhistg179}, and \ref{fig:qmc:nglhistg191} show the histogram of the global linear and nonlinear susceptibilities. It is to be noticed that we have less data points than in the case of the local susceptibilities. The resulting value of $d/z'$ is plotted in Fig.~\ref{fig:qmc:gldzgamma}.
From these results we confirm that the data for the global susceptibilities exhibit similar behavior to those for the local susceptibilities both in the histograms and the exponent $z'$. This is important because the experimental data from the D-Wave device are available only for the global susceptibilities.

\begin{figure}[tbp]
	\centering
    \includegraphics[width=.9\linewidth]{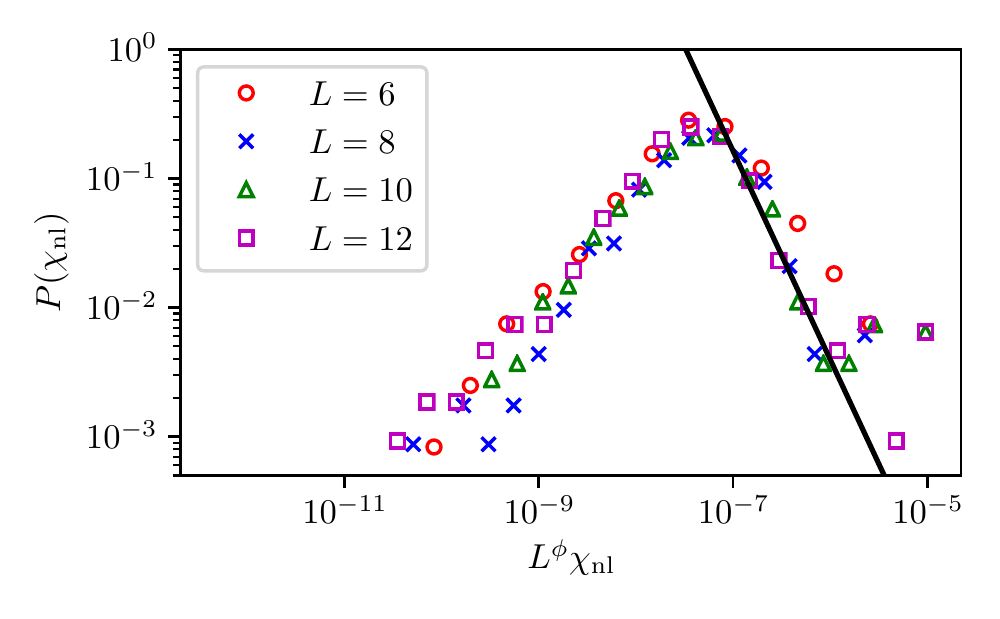}
    \caption{ $P(\chi_{\mathrm{nl}})$ which is similar to Fig.~\ref{fig:qmc:nlochistg1895} but for the global nonlinear susceptibility at $\Gamma = 1.865$. The black solid line shows a fit to the data whose slope is  $-1.616 \pm 0.25$, which indicates that $d/3z'$ is around $0.616 \pm 0.25$ or $d/z' \simeq 1.85 \pm 0.75$, which seems to be quite small compared to the data from global linear susceptibility in Fig.~\ref{fig:qmc:gldzgamma}.}
	\label{fig:qmc:nglhistg1865}
\end{figure}

Lastly, we point out that the data for the global nonlinear susceptibility at $\Gamma = 1.865$ as shown in Fig.~\ref{fig:qmc:nglhistg1865}
gives the value of the exponent $d/z'\simeq 1.3$ much smaller than the value indicated in Fig.~\ref{fig:qmc:gldzgamma}, around 4.  It may be due partly to insufficient statistics but further investigation is needed.

\section{Experiment on the D-Wave quantum annealer}
\label{sec:dwave}

We next carry out experiments of the Griffiths-McCoy singularity on the D-Wave Systems Inc.~2000Q  at NASA Ames Research Center.
\subsection{Method of experiment}
\begin{figure}[tbp]
	\centering
    \includegraphics[width=.9\linewidth]{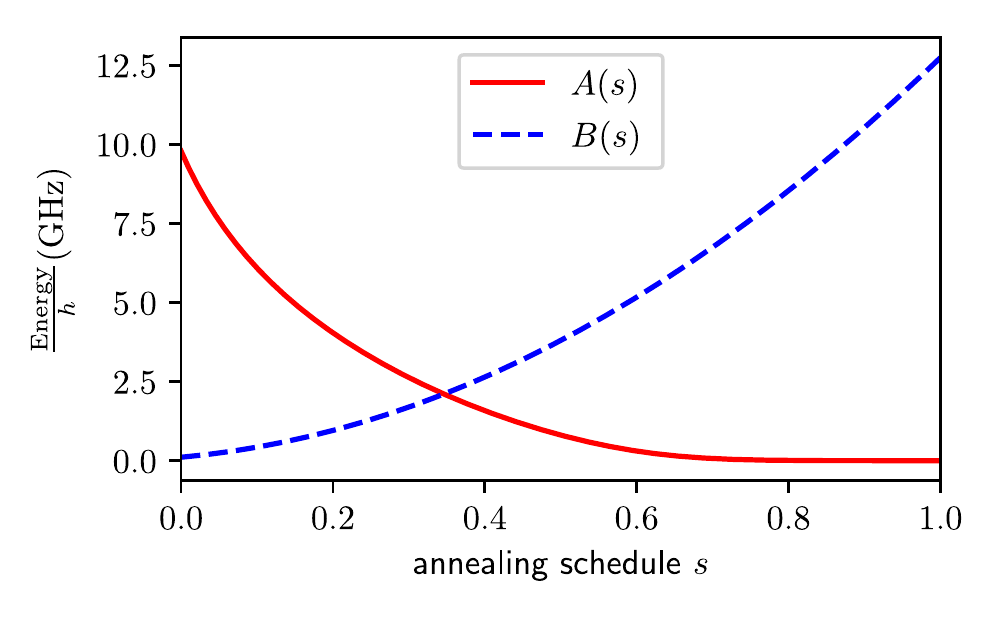}
    \caption{ The annealing schedules $A(s)$ and $B(s)$ on the D-Wave 2000Q.}
	\label{fig:dwave:AB}
\end{figure}

We follow the convention to write the Hamiltonian used in the D-Wave experiment as
\begin{align}
    \label{eq:dwave:H}
    H(s) =  - \frac{A(s)}{2}\left( \sum_{i}\hat{\sigma}_i^x \right) + \frac{B(s)}{2} \left(\sum_{\braket{i,j}} J_{ij}\hat{\sigma}_i^z \hat{\sigma}_j^z \right),
\end{align}
where $s$ is the time parameter running from 0 to 1, and the time dependence of $A(s)$ and $B(s)$ is depicted in Fig.~\ref{fig:dwave:AB}.
To generate a state as close as possible to a quantum thermal equilibrium state on the D-Wave machine, we employ the anneal-pause-quench protocol where we first perform an anneal up to value $s=s_{*}$, pause at this point for a while, then quench to $s=1$ as rapidly as possible \cite{King2018,Harris2018,King2019}. 
By applying this protocol, the D-Wave device may return the spin configuration $\v{\sigma}$ sampled from a distribution not far from the canonical ensemble with the Hamiltonian $H(s_{*})$ of Eq.~(\ref{eq:dwave:H}). The accuracy of this procedure nevertheless needs careful scrutiny as discussed below and in Refs.~ \cite{King2018,Harris2018,Izquierdo2020,Bando2020}.

We use the same diluted Chimera graph as in Sec. \ref{sec:qmc} for the experiment on the D-Wave machine. We choose the amplitude of interactions according to the probability distribution
\begin{align}
    P(J_{ij}) = \frac{1}{6}\sum_{k=0}^{5}\delta(J_{ij} + 0.1k).
    \label{eq:Prob_DW}
\end{align}
The unit of $J_{ij}$ here follows the convention of the D-Wave machine such that the maximum possible value is $|J_{ij}|=1$. To keep the largest $|J_{ij}|$ to be 0.5 in the above distribution function Eq.~(\ref{eq:Prob_DW}) is expected to reduce the effect of the noise in the D-Wave machine
since a large amplitude of interactions tends to amplify analog errors on the machine \cite{RHprivatecomm}.
Notice that that fact that the values of $|J_{ij}|$ in this Eq.~(\ref{eq:Prob_DW})  are a half of those for the QMC in Eq.~(\ref{eq:Jij_QMC}) is unimportant due to the difference in energy units used in both approaches.
The parameters used throughout the experiments are listed in Table \ref{tab:dwave:dwaveparam}.
\begin{table*}[ht]
    \centering
    \caption{Parameters for experiments on the D-Wave machine. $N_\mathrm{rand}$ is the number of random instances, $N_\mathrm{rand}'$ is the number of random instances for the histogram of susceptibility, $N_\mathrm{rep}$ is the number of annealing repetitions,  $L$ is the square root of the number of Chimera units ($8L^2$ is the total number of sites to be denoted as $N$),  $s_{* \mathrm{min}}$ is the minimum value of annealing schedule,  $s_{* \mathrm{max}}$ is the maximum value of the annealing schedule, $N_\Gamma$  is the number of transverse field values, $t_1$ is the anneal time in the anneal-pause-quench protocol, $t_2 - t_1$ is the pause time, $t_f-t_2$ is the quench time.
    }
    \label{tab:dwave:dwaveparam}
    \begin{tabular}{c c c c c c c c c c}
        \hline
        \hline
        $N_\mathrm{rand}$ & $N_\mathrm{rand}'$ &$N_\mathrm{rep}$ & $L$ & $s_{* \mathrm{min}}$ & $s_{* \mathrm{max}}$ &  $N_\Gamma$ & $t_1$ & $t_2 - t_1$ &  $t_f-t_2$ \\
        \hline
        255 & 438 & 100 & $8, 12, 16$ & 0.36 & 0.41 & 50 & 1000$s_{*}\,\mu$s & 100$\mu$s & $(1-s_{*})\,\mu$s\\
        \hline
    \end{tabular}
\end{table*}

First we describe how to measure the magnetization. For each annealing process, we obtain a set of classical values of spins $\{\sigma_1, \sigma_2, \cdots\}$ following the anneal-pause-quench protocol with longitudinal field switched off and store the magnetization,
\begin{align}
    m_a = \frac{1}{N}\sum_{i=1}^{N}\sigma_i.
\end{align}
This annealing process is repeated $N_{\mathrm{rep}}$ times to obtain a set of magnetization values $m_a$ $(1\leq a \leq N_{\mathrm{rep}})$. Given enough interval time (200$\mu $s) between consecutive annealing processes, we can assume that samples of the magnetization $m_a$ are uncorrelated with each other. The $n$th moment of the magnetization is calculated from these as
\begin{align}
    \braket{m^n} = \frac{1}{N_{\mathrm{rep}}}\sum_{a=1}^{N_{\mathrm{rep}}}m_a^n.
\end{align}
Various physical quantities such as the Binder ratio are derived from the above $n$th moment of magnetization.

To estimate the linear and nonlinear susceptibilities $\chi$ and $\chi_{\mathrm{nl}}$, we measure the magnetization as a function of the longitudinal field $h$.
Linear and nonlinear susceptibilities are obtained by applying a polynomial fit to the magnetization curve,
\begin{align}
    m \simeq \chi h - \chi_{\mathrm{nl}} h^3 + \cdots,
\end{align}
for each given instance of random interactions.

Due the analog nature of the D-Wave device, naive experiments without noise mitigation produce data with very limited reliability, in particular in the present case of the detection of a delicate phenomenon. We therefore apply two kinds of techniques,  calibration of individual flux bias and the standard gauge averaging, to reduce noise for more reliable results.  A technical description of the former method is given in Appendix \ref{sec:appendix_calibration}.

\subsection{Results}

We first analyze the data to determine the transition point and then estimate the exponent $d/z'$.

\subsubsection{Transition point}
\begin{figure*}[tbp]
	\centering
    \includegraphics[width=.4\linewidth]{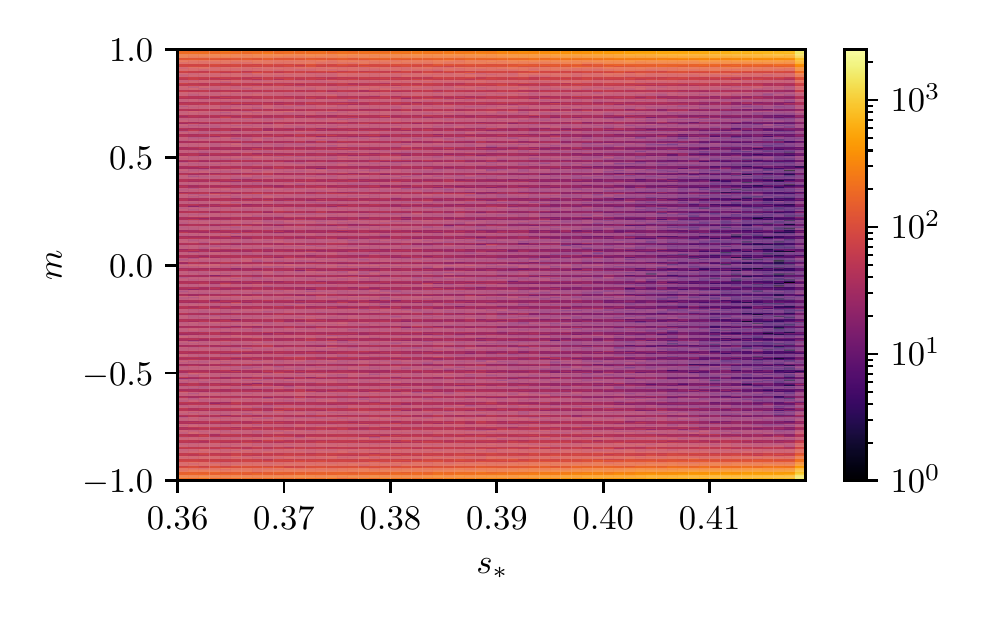}
    \includegraphics[width=.4\linewidth]{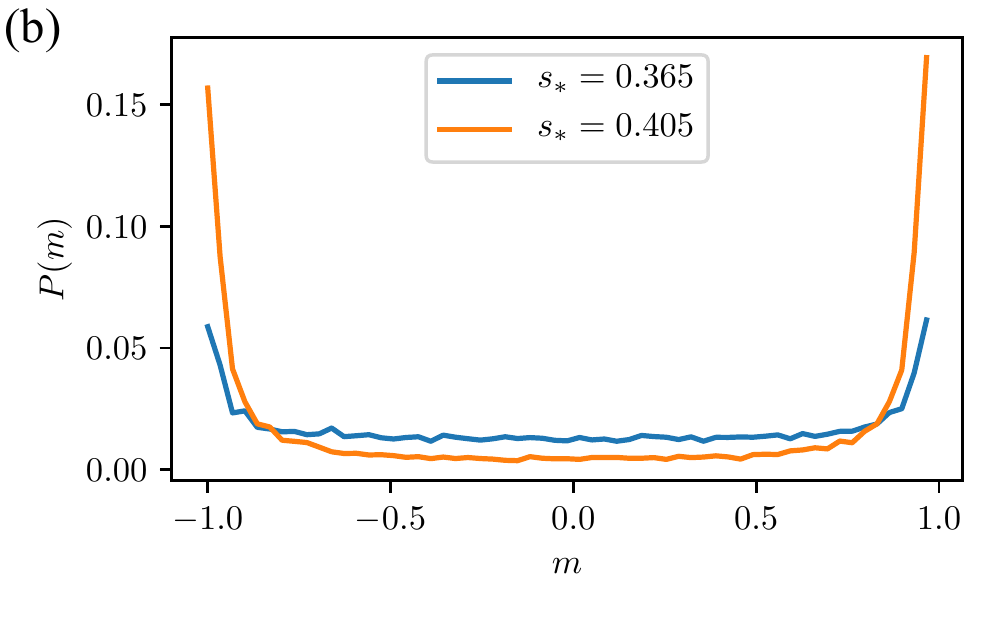}
    \includegraphics[width=.4\linewidth]{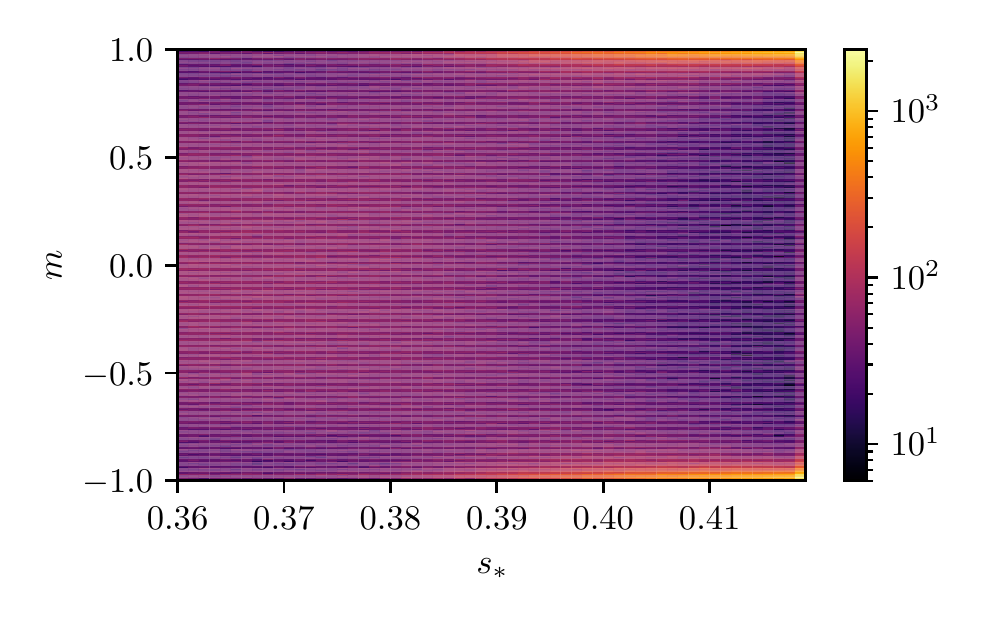}
    \includegraphics[width=.4\linewidth]{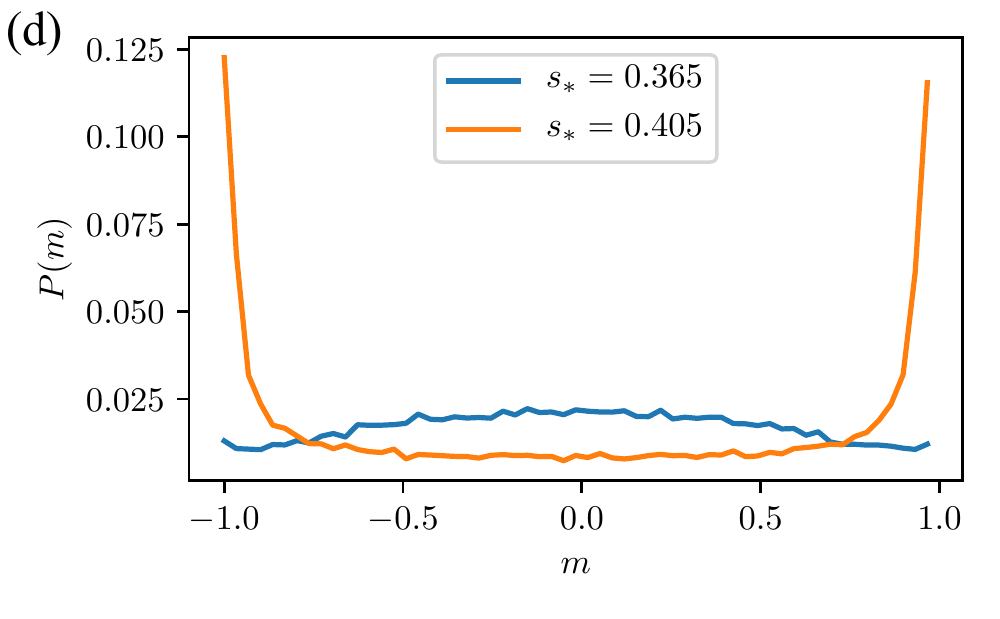}
    \includegraphics[width=.4\linewidth]{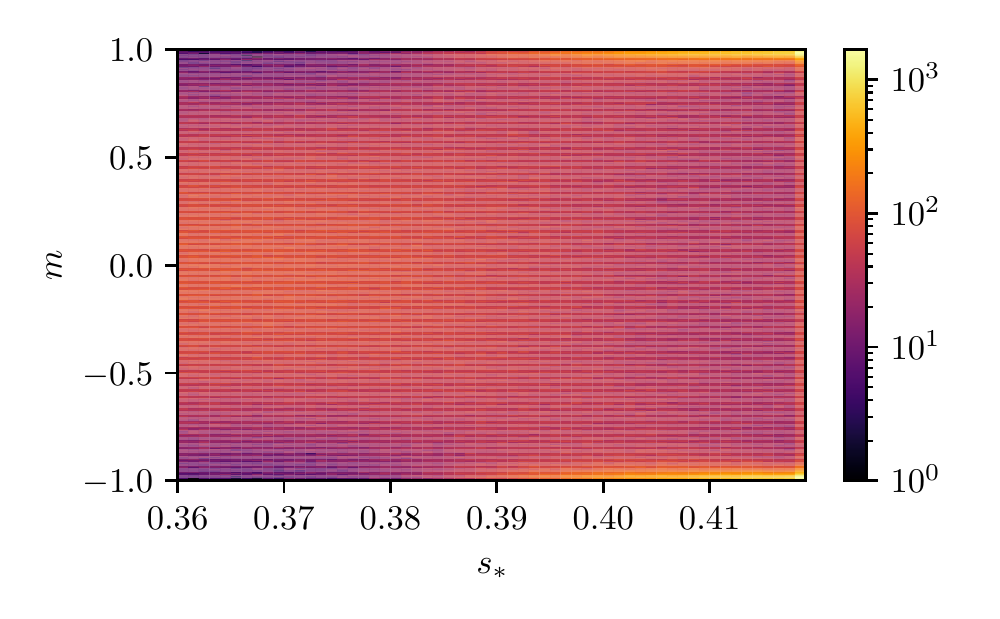}
    \includegraphics[width=.4\linewidth]{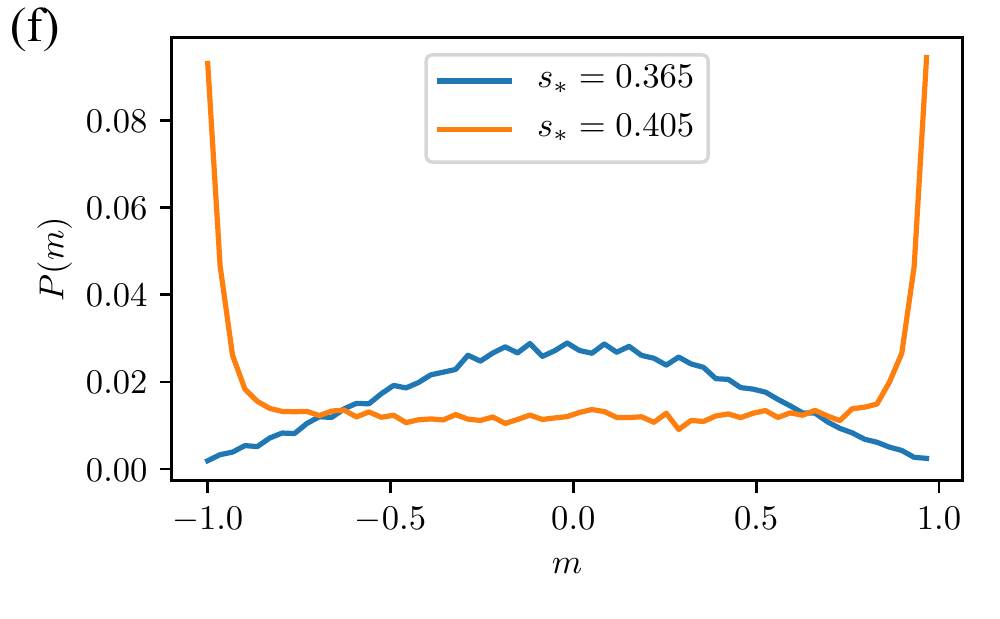}
    \caption{ (Left column) Histograms (heat map) of magnetization with system sizes (a) $L=8$, (c) $L=12$, and (e) $L=16$. The horizontal axis $s_{*}$ denotes the pause point during the anneal-pause-quench protocol, corresponding to a finite $\Gamma$ of the transverse field. The vertical axis is for the magnetization. Up to a certain point $s_{*} \simeq 0.39$, the magnetization tends to be distributed around zero, which implies that the system is in the paramagnetic phase. In the region with $s_{*}$ larger than 0.39, the magnetization tends to have values of saturation $\pm1$, and the system is in the ferromagnetic phase. Note that the color code is in logarithmic scale. (Right column) Cross sections of the heat map on the left column with (b) $L=8$, (d) $L=12$, and (f) $L=16$, respectively, at select values of $s_{*}$,  0.365 (in the presumed paramagnetic phase) and 0.405 (in the presumed ferromagnetic phase).}
	\label{fig:dwave:mhist}
\end{figure*}

Let us start with the distribution of the magnetization. Figure \ref{fig:dwave:mhist} shows the histogram of the magnetization as a function of the pause point $s_*$ for linear system sizes $L=8$, $12$, and $16$.
This figure, especially for the largest system with $L=16$,  shows that the magnetization is distributed around zero below a certain value of $s_{*}$ close to 0.39 and tend to have a peak near $\pm 1$ above this threshold point.
This change of the shape of the histogram suggests the existence of a phase transition at around $s_{*c} \simeq 0.39$. Also, especially for the small system size, we observe that some samples have large values close to $\pm 1$ even in the region supposed to be the paramagnetic phase. One of the possible reasons for this unexpected behavior is a systematic error coming from the quench process during the anneal-pause-quench protocol \cite{King2018,Izquierdo2020}: At the end of the protocol, the annealing parameter $s$ is changed from $s_{*}$ to $1$ as quickly as possible, which corresponds to a ``switching-off" of the transverse field. Although this quenching process is supposed to be performed quickly, the elapsed time in this process ($\simeq 1\mu$s) may still be long enough to affect the final state of the system, leading to a broad distribution of the magnetization in the presumed paramagnetic phase, which should not be the case in theory.
The data for the small size $L=6$ in the paramagnetic phase are strongly affected by this imperfection and we should take sufficient care in the analysis.

\subsubsection*{Binder ratio and averaged magnetization}

\begin{figure}[tbp]
	\centering
    \includegraphics[width=.7\linewidth]{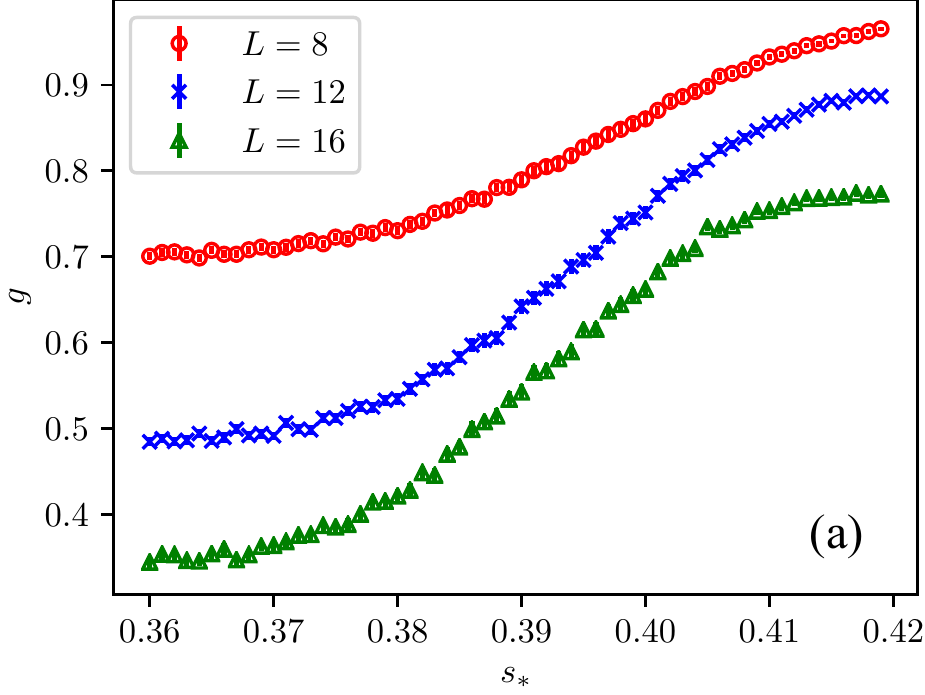}
    \includegraphics[width=.7\linewidth]{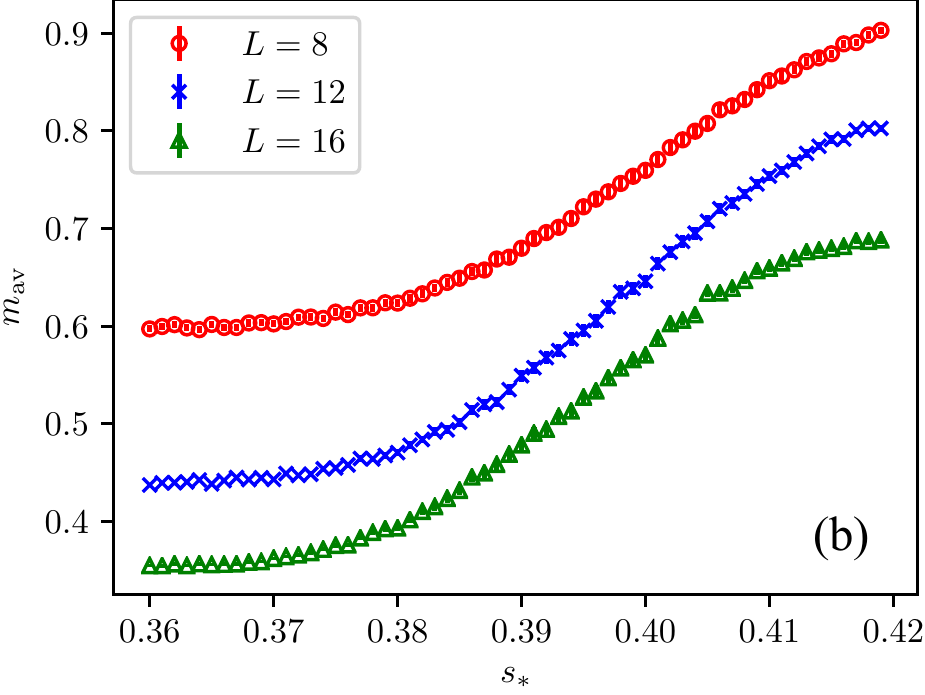}
    \caption{ (a) The Binder ratio for three linear sizes from the D-Wave experiment. (b) Averaged magnetization for three linear sizes obtained from the D-Wave experiment.}
	\label{fig:dwave:Binderandmag}
\end{figure}

The results of the Binder ratio and the averaged magnetization $m_{\mathrm{av}} = [\braket{|m|}]$
are shown in Fig.~\ref{fig:dwave:Binderandmag}. 
We usually expect the Binder ratio with different system sizes to cross at the transition point as in Fig.~\ref{fig:qmc:Binder}. However, in the present case, we find no crossing point in Fig.~\ref{fig:dwave:Binderandmag}, and both the Binder ratio and the averaged magnetization have finite values even in the paramagnetic region, most prominently for the small system size. This behavior can be understood in the same way as in the case of the histogram of the magnetization: The relatively slow quenching process in the anneal-pause-quench protocol may allow the system to follow the decrease of the transverse field, driving the system toward ferromagnetic ordering. The large values of the Binder ratio and magnetization at small $s_{*}$, especially for $L=6$, should reflect the broad distribution of magnetization seen in Fig.~\ref{fig:dwave:mhist}. Nevertheless, we at least find that the magnetization for the largest system $L=16$ is likely to have an inflection point around $s_{*} \simeq 0.39$, remotely suggesting the existence of a phase transition point around this value.

\subsubsection*{Global linear and nonlinear susceptibilities}

\begin{figure}[tbp]
	\centering
    \includegraphics[width=.7\linewidth]{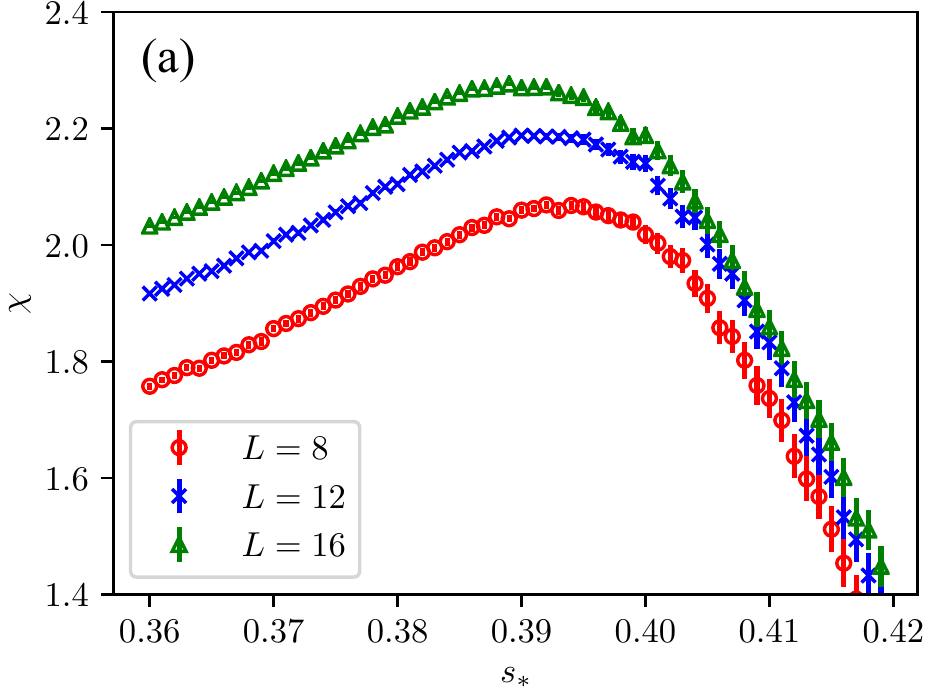}
    \includegraphics[width=.7\linewidth]{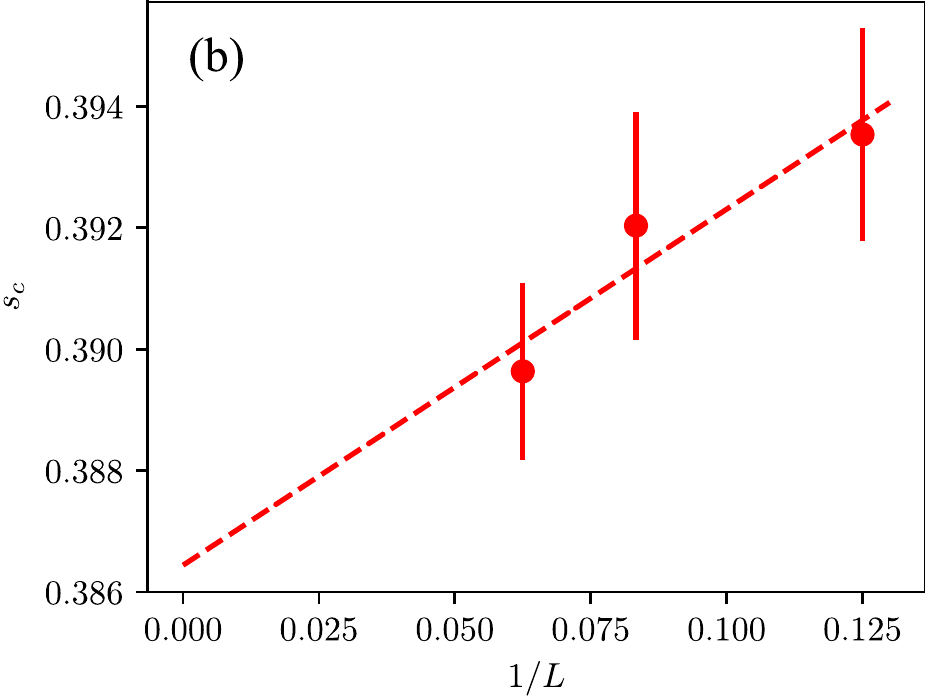}
    \caption{ (a) D-Wave data for the global linear susceptibility for system sizes $L=8, 12$ and 16. (b) Position of peaks as a function of the inverse system size $1/L$. The red dashed line shows a linear fit.} 
	\label{fig:dwave:susp}
\end{figure}

Figure \ref{fig:dwave:susp}(a) shows the global linear susceptibility $\chi$ for three system sizes.
We find that there are peaks at around $s_{*} \simeq 0.39$ and the position becomes smaller as the system size increases. Extrapolation to the infinite-size limit as shown in Fig.~\ref{fig:dwave:susp} gives $s_c = 0.386 \pm 0.002$, which is consistent with the data of the histogram of the magnetization in Fig.~\ref{fig:dwave:mhist}. Since the data, especially for the smallest size $L=8$, are not necessarily very reliable, we choose not to go beyond the estimate of the approximate value of the transition point and avoid finite-size scaling analysis for critical exponents.

\begin{figure}[tbp]
	\centering
    \includegraphics[width=.7\linewidth]{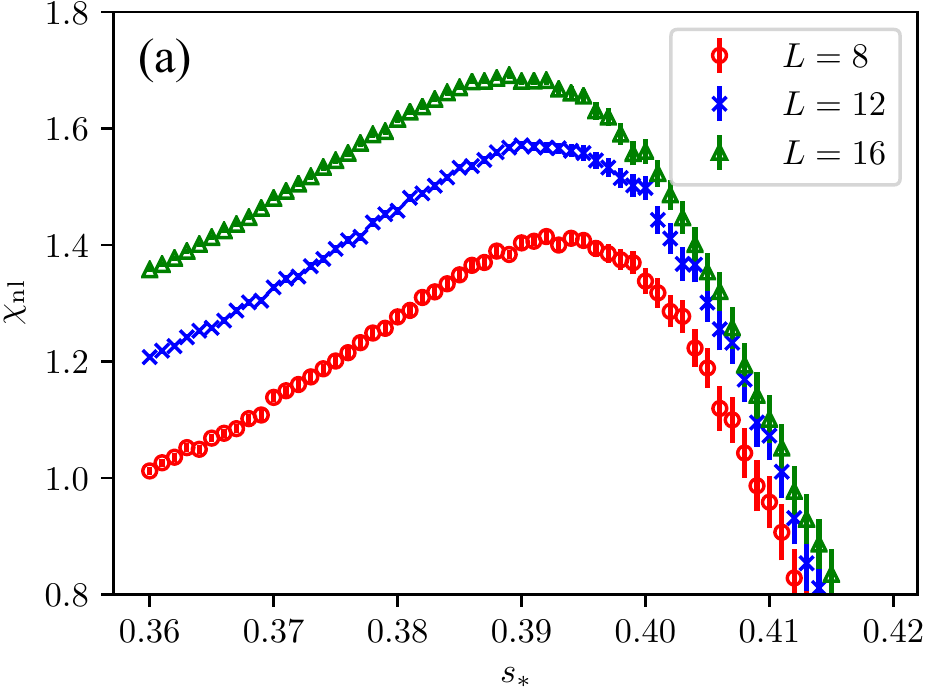}
    \includegraphics[width=.7\linewidth]{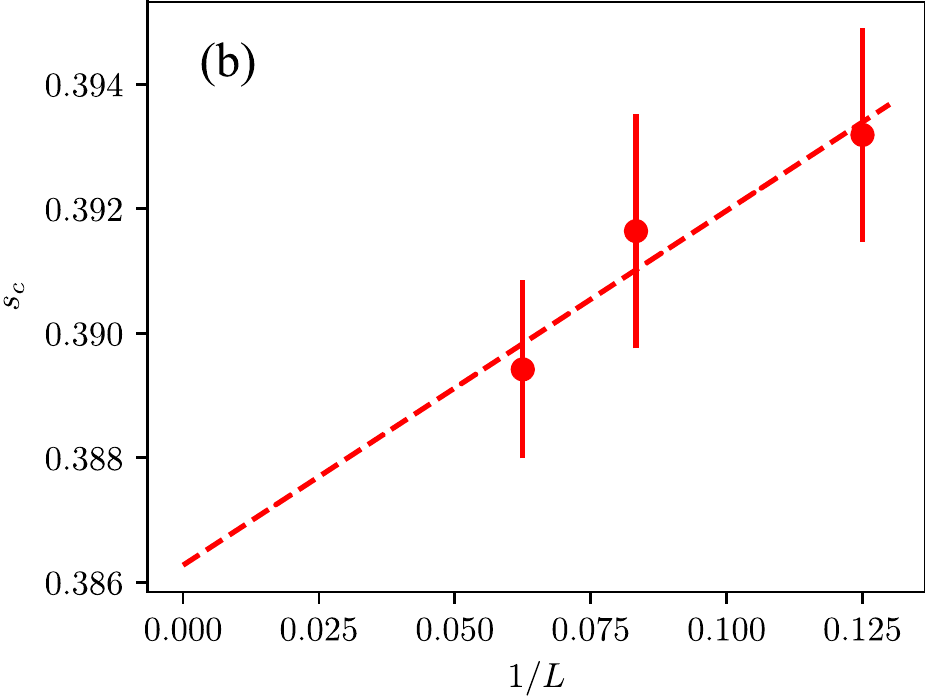}
    \caption{ (a) D-Wave data for the global nonlinear susceptibility. The aspect ratio of this graph is the same as in Fig.~\ref{fig:dwave:susp} for direct comparison. (b) Peak position of the nonlinear susceptibility as a function of the inverse system size $1/L$.}
	\label{fig:dwave:nlsusp}
\end{figure}

We also measured the global nonlinear susceptibility as shown in Fig.~\ref{fig:dwave:nlsusp}(a), where the aspect ratio is the same as in Fig.~\ref{fig:dwave:susp}, for convenience of comparison.
Although the peak is a little bit narrower compared to the global linear susceptibility, the peak position is almost the same as in the global linear susceptibility, leading to the same critical point $s_c = 0.386 \pm 0.002$ in the infinite system size limit as shown in Fig.~\ref{fig:dwave:nlsusp}(b).

\subsubsection{Estimation of the exponent}

\begin{figure}[tbp]
	\centering
    \includegraphics[width=.9\linewidth]{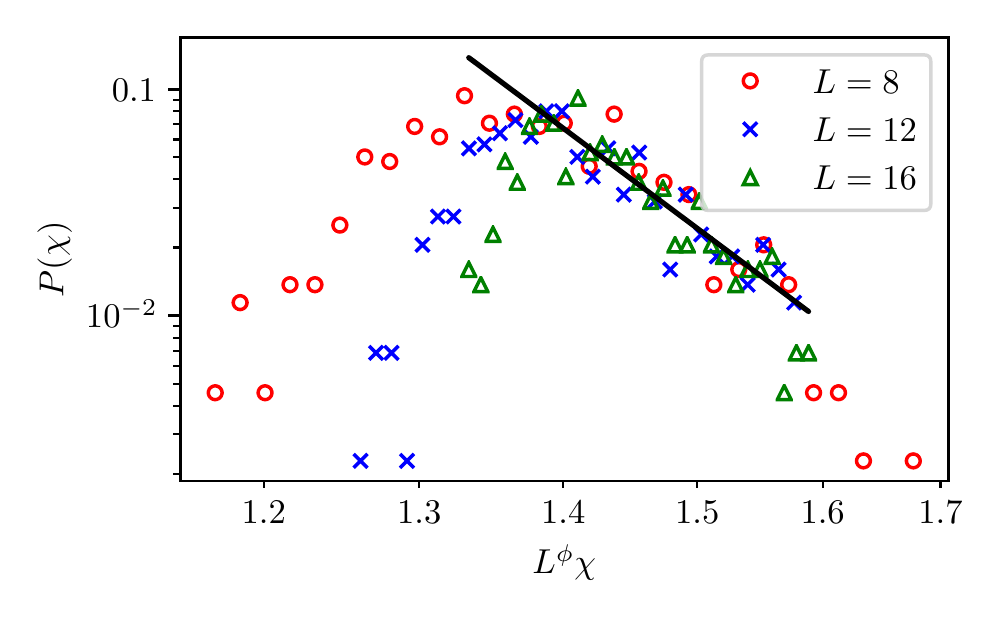}
    \caption{ Log-log plot of the D-Wave data for $P(\chi)$ of the global linear susceptibility $\chi$ at the pause point $s_{*} = 0.365$ (far from the critical point).
The black solid line is a fit to a line of slope $-14.7\pm 1.74$, implying that $d/z'\simeq 13.7\pm 1.7$.}
	\label{fig:dwave:glhists0365}
\end{figure}

We now  analyze the histogram of susceptibilities to estimate the exponent $d/z'$. Figure \ref{fig:dwave:glhists0365} shows the histogram of the global linear susceptibility with the pause point $s_{*} = 0.365$ which is far from the critical point and is expected to be in the paramagnetic phase. Notice that when applying a linear fit to estimate $d/z'$, we exclude the data with $P(\chi) \leq 10^{-2}$ partly because these data may not be reliable due to insufficient statistics. Another reason is given below.
We find that a linear fit to the data does not seem too bad in the region of large susceptibility, especially for the largest system size. We also observe a long tail of small susceptibilities at the left part of the graph, notably for the small size.  This tail may be understood by considering the feature of the histogram of the magnetization in Fig.~\ref{fig:dwave:mhist}, where samples with values near saturation $\simeq 1$ exist even in the paramagnetic region. These samples may have small susceptibilities because the system does not respond to the field when the magnetization is close to saturation, resulting in the long tail of small susceptibility at the left part of Fig.~\ref{fig:dwave:glhists0365}. 
Samples with large negative magnetization close to $-1$ would respond very strongly to the positive field $h>0$, flipping the state from $m\simeq -1$ to $m\simeq 1$, and the tail of the distribution for very large $\chi$ would correspond to such cases.  We therefore drop the data in the right-most tail of distribution from the analysis.
In contrast, samples with small magnetization in the paramagnetic phase are likely to have reasonable properties, which would yield the moderately large susceptibility compared to the case with saturated magnetization.  It therefore seems reasonable to consider that analyses of data with moderately large susceptibilities would give relatively reliable results.
We then apply a linear fit to the data with large, but not too large, values of the susceptibility.  We find that $d/z'$ is around 13.7 for the present pause point $s_{*} = 0.365$.

\begin{figure}[tbp]
	\centering
    \includegraphics[width=.9\linewidth]{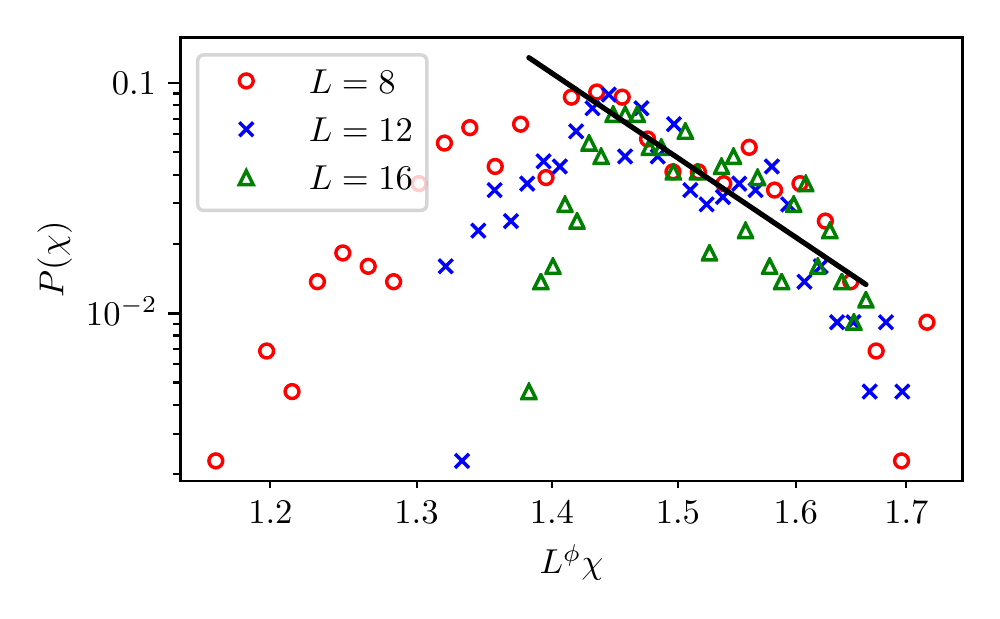}
    \caption{ Log-log plot of the D-Wave data for the histogram $P(\chi)$ of the global linear susceptibility $\chi$ with the pause point $s_{*} = 0.369$. The black solid line shows a fit to data whose slope is $-12.3\pm 2.0$, which indicates that the exponent $d/z'$ is around 11.3$\pm 2.0$.}
	\label{fig:dwave:glhists0369}
\end{figure}

\begin{figure}[tbp]
	\centering
    \includegraphics[width=.9\linewidth]{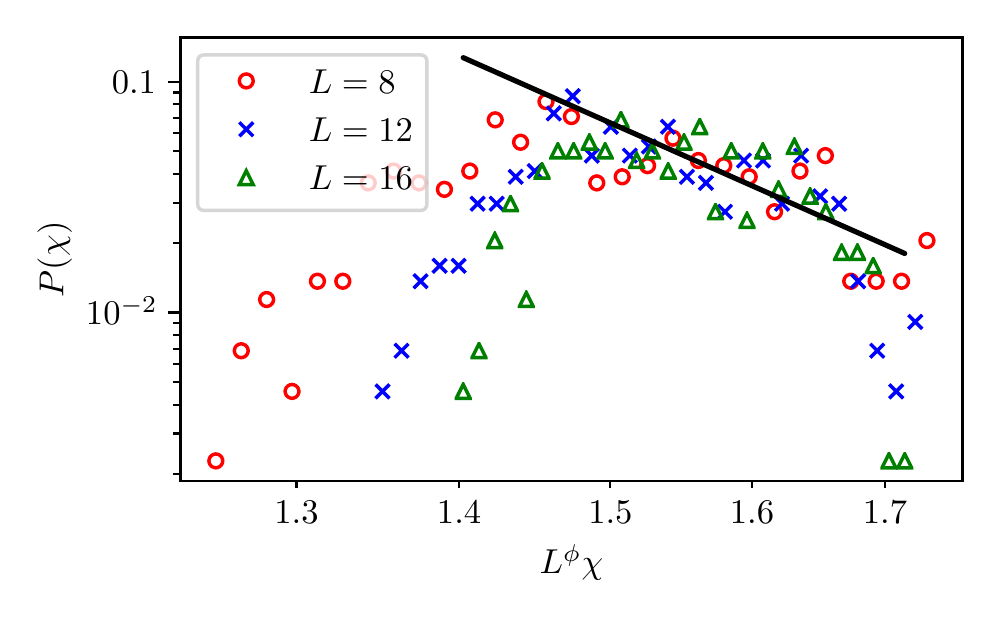}
    \caption{ Log-log plot of the D-Wave data for the histogram $P(\chi)$ of the global linear susceptibility $\chi$ with the pause point $s_{*} = 0.375$. The black solid line shows a fit to data whose slope is $-9.7\pm 2.1$, which indicates that the exponent $d/z'$ is around $8.7\pm 2.1$.}
	\label{fig:dwave:glhists0375}
\end{figure}

\begin{figure}[tbp]
	\centering
    \includegraphics[width=.9\linewidth]{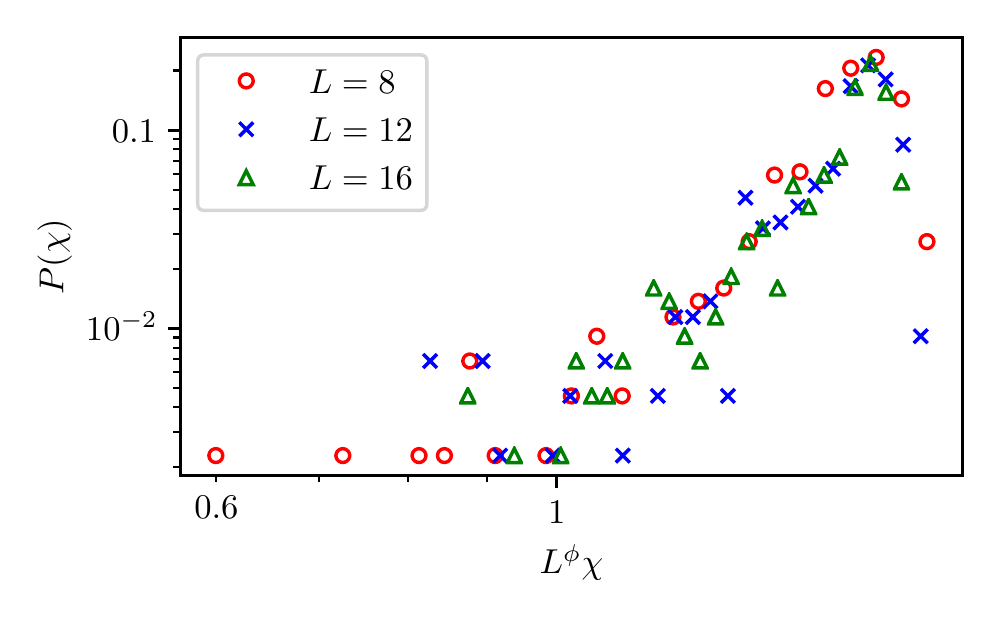}
    \caption{ Log-log plot of the D-Wave data for the histogram $P(\chi)$ of the global linear susceptibility $\chi$ with the pause point $s_{*} = 0.394$ (above the critical point). No power-law decay behavior is observed on this figure and $P(\chi)$ roughly grows monotonically as the global linear susceptibility increases.}
	\label{fig:dwave:glhists0394}
\end{figure}

The distributions $P(\chi)$ at the pause points $s_{*} = 0.369$ and $0.375$, which are closer to the critical point but still in the paramagnetic phase, are shown in Figs. \ref{fig:dwave:glhists0369} and \ref{fig:dwave:glhists0375}, respectively.  The resulting exponents are $d/z'\simeq 11.3\pm 2.0$ and $8.7\pm 2.1$ for $s_{*} = 0.369$ and $0.375$, respectively.
 In contrast, the data for larger $s_{*}$ are difficult to analyze to extract the exponent reliably. The data for the region above the critical point, $s_{*} = 0.394$, is shown on Fig.~\ref{fig:dwave:glhists0394}.
No simple power-law decay is observed here.  We find that $P(\chi)$ monotonically increases as $\chi$ increases. The same behavior is observed for the local linear susceptibility obtained by quantum Monte Carlo as shown in Fig.~\ref{fig:qmc:lochistg14}.

\begin{figure}[tbp]
	\centering
    \includegraphics[width=.9\linewidth]{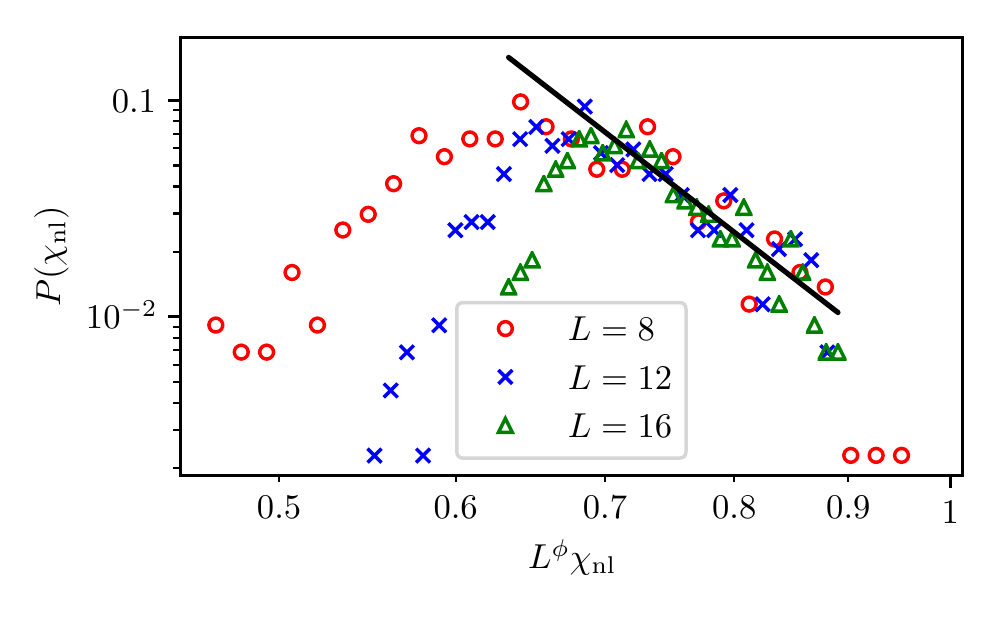}
    \caption{ Log-log plot of the D-Wave data for the global nonlinear susceptibility $\chi_{\mathrm{nl}}$ at the pause point $s_{*} = 0.365$ (far from the critical point). The black solid line is a fit to data with slope $-8.0\pm 1.1$, i.e., $d/3z'$ is around $7.0\pm 1.1$.}
	\label{fig:dwave:nglhists0365}
\end{figure}

\begin{figure}[tbp]
	\centering
    \includegraphics[width=.9\linewidth]{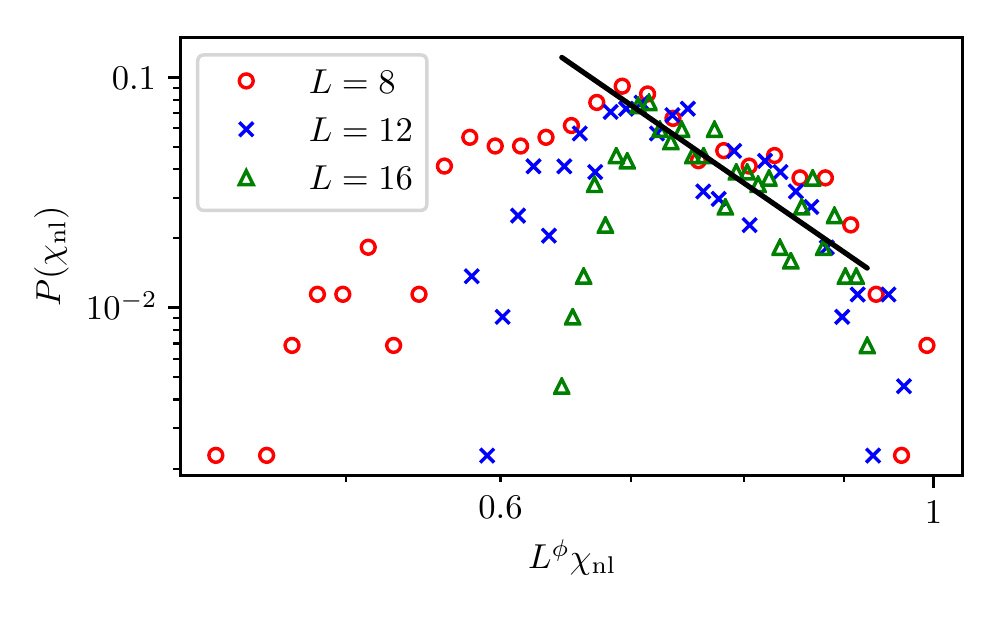}
    \caption{ Log-log plot of the D-Wave data for $P(\chi_{\mathrm{nl}})$ of the global nonlinear susceptibility obtained at the pause point $s_{*} = 0.369$ (far from the critical point). The black solid line shows a fit to data whose slope is $-5.8\pm 0.9$, which indicates that the exponent $d/3z'$ is around $4.8\pm 0.9$.}
	\label{fig:dwave:nglhists0369}
\end{figure}

Similar behavior is observed in the global nonlinear susceptibility. Figure \ref{fig:dwave:nglhists0365} shows the histogram $P(\chi_{\mathrm{nl}})$ for $s_{*} = 0.365$.
The histogram of the global nonlinear susceptibility with the pause point $s_{*} = 0.369$ is shown in Fig.~\ref{fig:dwave:nglhists0369}. 
We observe that $d/3z'$ decreases monotonically as in the global linear susceptibility, about $7.0\pm 1.1$ ($s_{*} = 0.365$) and $4.8\pm 0.9$ ($s_{*} = 0.369$).

\begin{figure}[tbp]
	\centering
    \includegraphics[width=.9\linewidth]{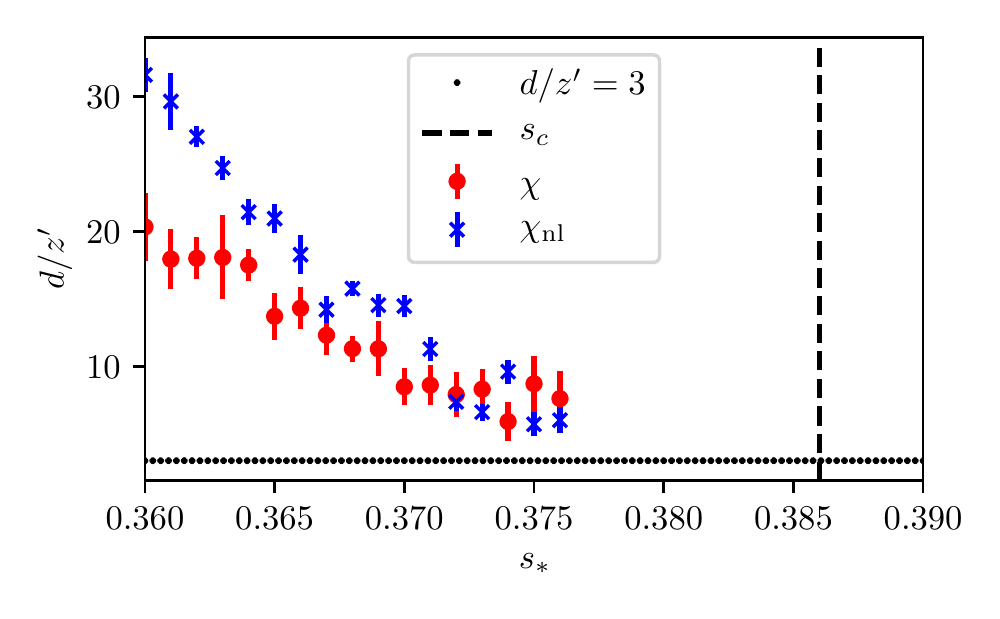}
    \caption{ The exponent $d/z'$ as a function of the pause point $s_{*}$ estimated from the linear and nonlinear susceptibilities. The vertical dashed line represents the critical point $s_c \simeq 0.386$ and the horizontal dotted line is for $d/z' = 3$, where the nonlinear susceptibility starts to diverge. 
    } 
	\label{fig:dwave:dzs}
\end{figure}

The relation between the exponent $d/z'$ and the pause point $s_{*}$ is shown in Fig.~\ref{fig:dwave:dzs}. 
Although data points for $s_{*}$ beyond 0.375 are excluded because of low reliability of the estimation of the slope of the histogram, the tendency seems consistent with the assumption that the inequality $d/z' < 3$ for the divergence of the nonlinear susceptibility is satisfied within the paramagnetic phase.

\subsection{Comparison of transition points from quantum Monte Carlo and D-Wave experiment}

\begin{figure}[tbp]
	\centering
    \includegraphics[width=.9\linewidth]{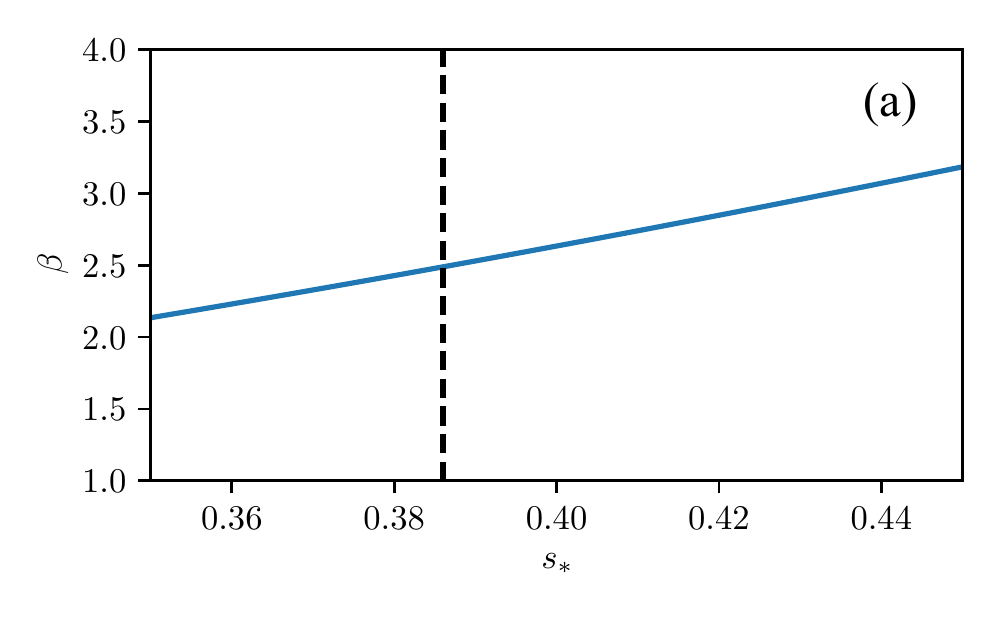}
    \includegraphics[width=.9\linewidth]{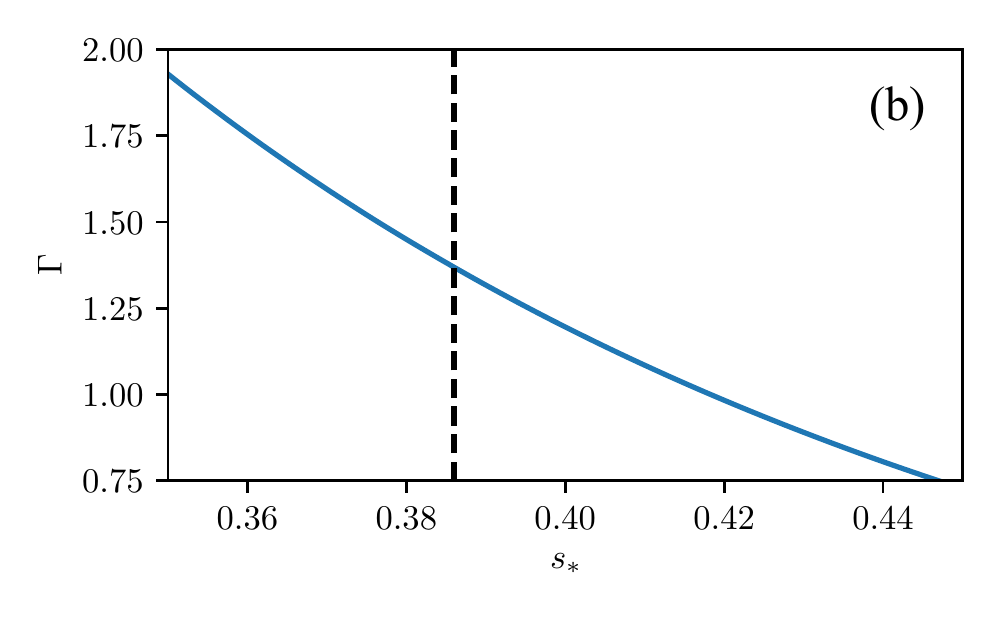}
    \caption{The relation between the pause point $s_{*}$ and the parameters (a) $\beta$ and (b) $\Gamma$. The black dashed line shows the point $s_{*} = s_c = 0.386$ at which $\beta = 2.49$ and $\Gamma = 1.37$.}
	\label{fig:dwave:ABGbeta}
\end{figure}

To confirm that the transition point estimated in Figs. \ref{fig:dwave:susp} and \ref{fig:dwave:nlsusp} is consistent with the result of the quantum Monte Carlo simulation, we relate the pause point $s_{*}$ and the pair of the inverse temperature and the transverse field $(\beta, \Gamma)$ by comparing the exponents of the Boltzmann factors as 
\begin{align}
    \label{eq:dwave:exprel}
    \exp{\left[-\beta_{\mathrm{phys}} H(s)\right]} \simeq \exp{\left[-\beta H\right]},
\end{align}
where $\beta_{\mathrm{phys}} \simeq 12$mK is the physical temperature of the D-Wave chip and $H(s)$ and $H$ denote the Hamiltonian of the D-Wave device and the quantum Monte Carlo in Eqs.~(\ref{eq:dwave:H}) and (\ref{eq:TFIM}), respectively. Assuming that the two exponentials in Eq.~(\ref{eq:dwave:exprel}) coincide, we obtain the relations between $s_{*}$ and $(\beta, \Gamma)$ as follows,
\begin{align}
    \label{eq:dwave:beta}
    \beta &= \frac{\beta_{\mathrm{phys}}B(s_{*})}{4}, \\
    \label{eq:dwave:G}
    \Gamma &= \frac{2 A(s_{*})}{B(s_{*})},
\end{align}
or with physical units written explicitly,
\begin{align}
    \label{eq:dwave:betaU}
    \beta &= \frac{1}{k_{B}\ \mathrm{J\cdot K^{-1}} \times 12 \mathrm{mK}}\frac{B(s_{*})\ \mathrm{GHz} \times h\ \mathrm{J\cdot s}}{4}, \\
    \label{eq:dwave:GU}
    \Gamma &= \frac{2 A(s_{*})\ \mathrm{GHz} \times h\ \mathrm{J\cdot s}}{B(s_{*})\ \mathrm{GHz} \times h\ \mathrm{J\cdot s}},
\end{align}
where $h$ and $k_{B}$ denote the Planck constant and the Boltzmann constant, respectively, the former not to be confused with the external field. Figure \ref{fig:dwave:ABGbeta} shows the relations in Eqs.~(\ref{eq:dwave:betaU}) and (\ref{eq:dwave:GU}).
From this figure we read that the critical point $s_{*} = 0.386$ corresponds to the parameter $\beta=2.49$ and $\Gamma = 1.37$ in the quantum Monte Carlo method, which is not far from $\Gamma \simeq 1.6$ with the temperature $\beta = 2.49$ $(T = 1/\beta \simeq 0.4)$ according to Fig.~ \ref{fig:qmc:extrapolate}(a). Although perfect quantitative agreement has not been expected, we have reached a reasonable degree of agreement.

\section{Discussion}
\label{sec:discussion}

We have carried out quantum Monte Carlo simulations and experiments on the D-Wave quantum annealer in order to investigate if the Griffiths-McCoy singularity is observed in the transverse-field Ising model with random ferromagnetic interactions on the diluted Chimera graph.  The results of quantum Monte Carlo indicate it to be very likely that there exists a parameter range within the paramagnetic phase where the local and global {\em nonlinear} susceptibilities diverge, implying the existence of the Griffiths-McCoy singularity. It is difficult to determine with confidence from our data whether or not the local and global {\em linear} susceptibilities diverge in the paramagnetic phase although it can well be the case if the present model belongs to the same universality class as the transverse-field Ising model on the square lattice with uniformly random ferromagnetic interactions and uniformly random transverse field \cite{Pich1998}.  

The data from the D-Wave device include a larger amount of uncertainties than those of quantum Monte Carlo due to the systematic bias in flux qubits and other sources of errors that are intrinsic to the analog quantum device.  In particular, the distribution of magnetization has a significant amount of data points near saturation $\pm 1$ within the paramagnetic phase even after careful calibrations to cancel the ferromagnetic bias.  Nevertheless, the moderately-large-value part of the distribution of linear and nonlinear susceptibilities can be considered fairly robust against the bias and errors because a state with saturated magnetization $\simeq 1$ has no room of further change toward larger values of magnetization and therefore will respond only weakly to the external field, contributing very little to the large-value part of the distribution of susceptibility. 
The very-large-value tail of the distribution may also be discarded for a similar reason and for insufficient statistics.
With this observation in mind, we suppose that the exponent $d/z'$ estimated from the part of moderately large values of susceptibility is relatively reliable.  In this way, we may conclude that the data can be interpreted to be consistent with the statement that $d/z'<3$ is satisfied in the paramagnetic phase, meaning the existence of the Griffiths-McCoy singularity. If this is indeed the case, the present study is the first case in which this very subtle phenomenon involving rare regions of ferromagnetic clusters, enhanced by quantum effects, has been observed in an analog quantum simulator. Further developments in device technologies and tools of data analyses will lead to improved reliability, and possibly promoting research activities toward quantum simulation of complex many-body systems by analog quantum simulators.

\begin{acknowledgments}

We thank Firas Hamze for stimulating discussions and Andrew King for
useful comments.  K.N.~would like to thank D-Wave systems, Inc.~and the
members of the company, especially Richard Harris, for giving advice on
the analysis using the D-Wave machine.  The research is based upon work
partially supported by the Office of the Director of National
Intelligence (ODNI), Intelligence Advanced Research Projects Activity
(IARPA) and the Defense Advanced Research Projects Agency (DARPA), via
the U.S. Army Research Office contract W911NF-17-C-0050. The views and
conclusions contained herein are those of the authors and should not be
interpreted as necessarily representing the official policies or
endorsements, either expressed or implied, of the ODNI, IARPA, DARPA, or
the U.S. Government. The U.S. Government is authorized to reproduce and
distribute reprints for Governmental purposes notwithstanding any
copyright annotation thereon.  The research of K.N. was supported by
JSPS KAKENHI Grant Number JP18J13685.

\end{acknowledgments}

\appendix
\section{Critical exponents}
\label{sec:appendix_exponents}
This appendix discusses the estimation of critical exponents $\beta, \gamma$, and $z$ at the ferro-para transition point from the data of quantum Monte Carlo. 

\begin{figure}[tbp]
	\centering
    \includegraphics[width=.9\linewidth]{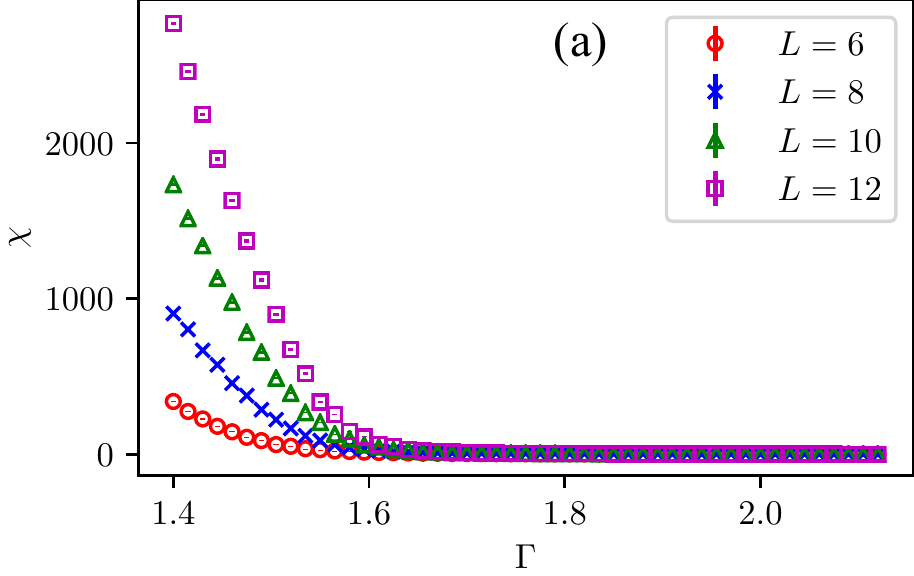}
    \includegraphics[width=.9\linewidth]{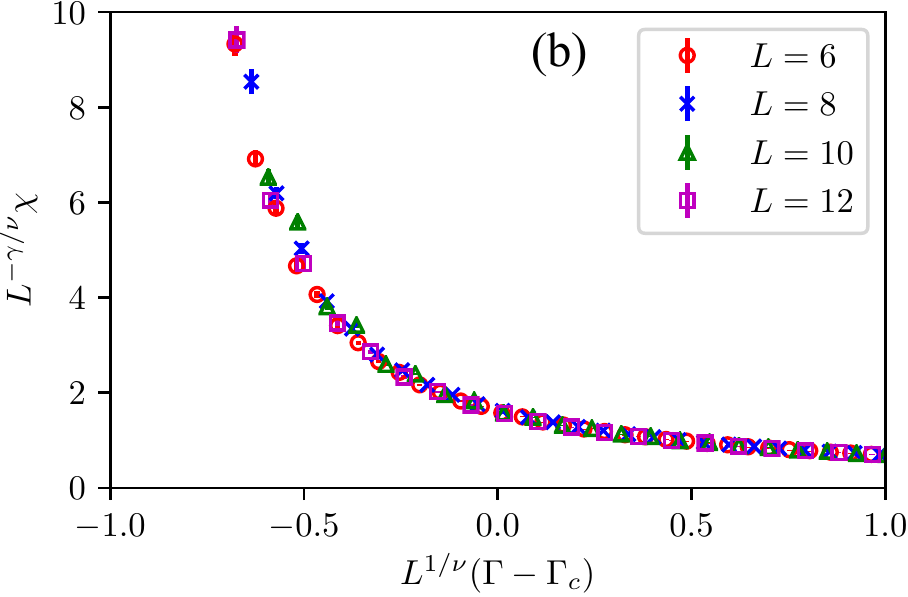}
    \caption{ (a) Global susceptibility $\chi$ with different system sizes $L=6, 8, 10, 12$ with the inverse temperature $\beta=20$. (b) Finite-size scaling analysis of the global susceptibility with $\gamma = 0.96(6)$.}
	\label{fig:qmc:susp}
\end{figure}

\begin{figure}[tbp]
	\centering
    \includegraphics[width=.9\linewidth]{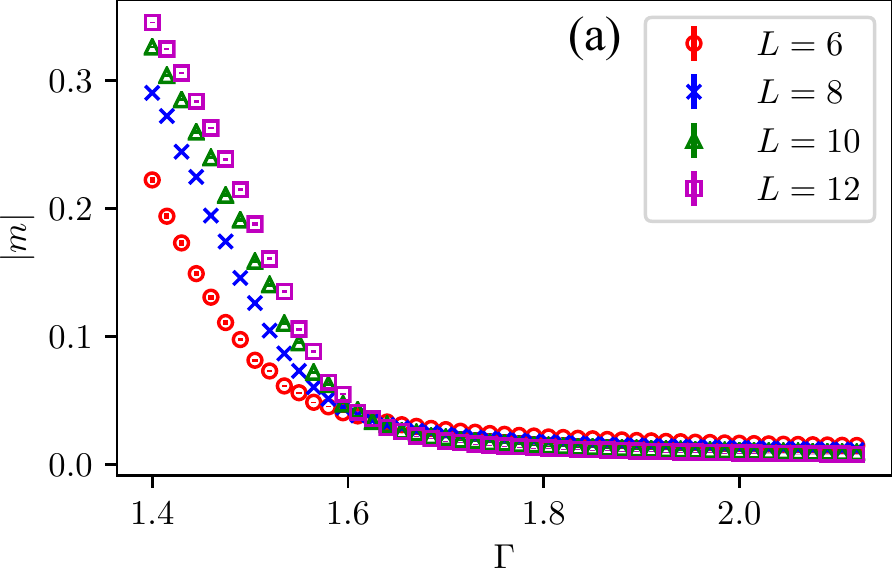}
    \includegraphics[width=.9\linewidth]{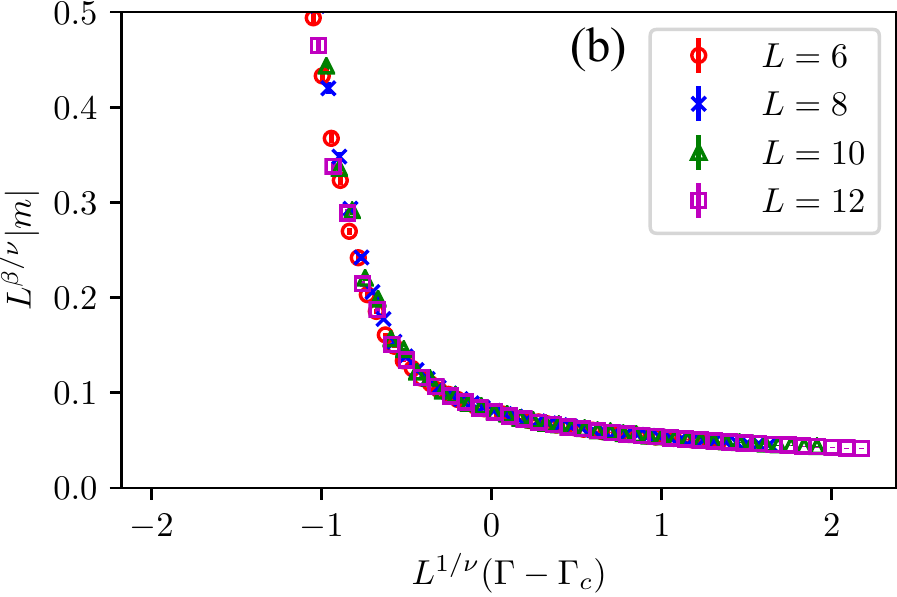}
    \caption{ (a) Magnetization $|m|$ with different system sizes $L=6, 8, 10, 12$ with the inverse temperature $\beta=20$. (b) Finite-size scaling analysis of the global susceptibility with the critical exponent $\beta = 0.95(6)$ for magnetization, which is not to be confused with the inverse temperature.}
	\label{fig:qmc:mag}
\end{figure}

\begin{figure}[tbp]
	\centering
    \includegraphics[width=.9\linewidth]{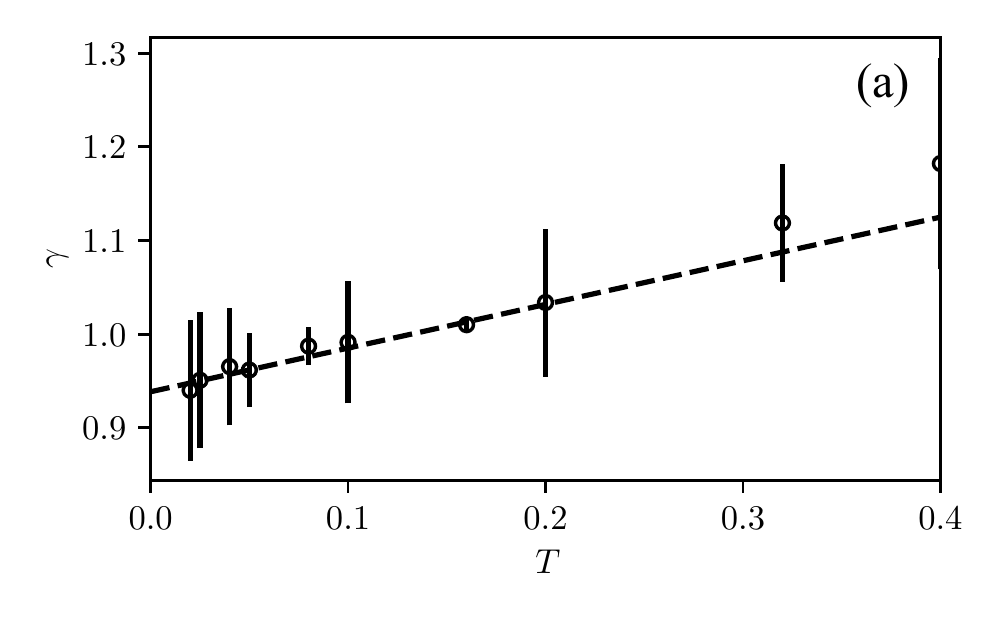}
    \includegraphics[width=.9\linewidth]{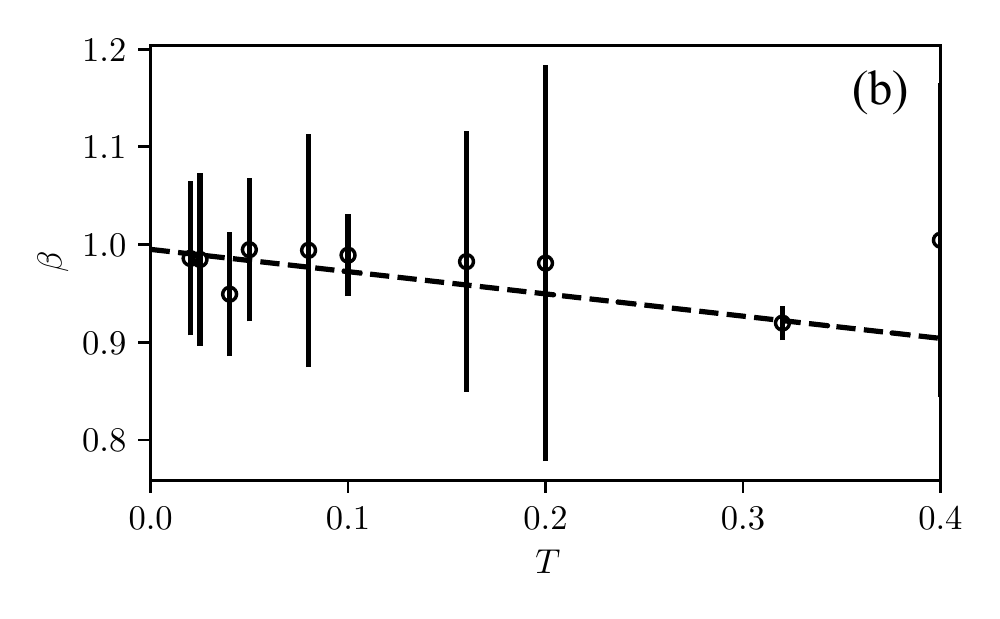}
    \caption{Critical exponents of the global susceptibility $\gamma$ in (a) and the magnetization $|m|$ in (b) as functions of the temperature $T$. Black dashed lines represent linear fitting of the data.}
	\label{fig:qmc:extrapolate2}
\end{figure}

We have carried out finite-size scaling analyses of the global susceptibility $\chi$ and the magnetization $|m|$ as in Figs. \ref{fig:qmc:susp} and \ref{fig:qmc:mag}.  Estimated values of the exponent $\nu$ and the transition point $\Gamma_{\rm c}$ described in Sec. \ref{sec:QMC_result} give a reasonable collapse of the data. The finite-temperature values of the exponents $\gamma$ and $\beta$ are extrapolated to zero temperature as in Fig.~\ref{fig:qmc:extrapolate2}.  Although the extrapolation suggests $\gamma$ to be close to 0.94 and $\beta$ to 1.0, these figures show large uncertainties in these estimates.

\begin{figure}[tbp]
	\centering
    \includegraphics[width=.8\linewidth]{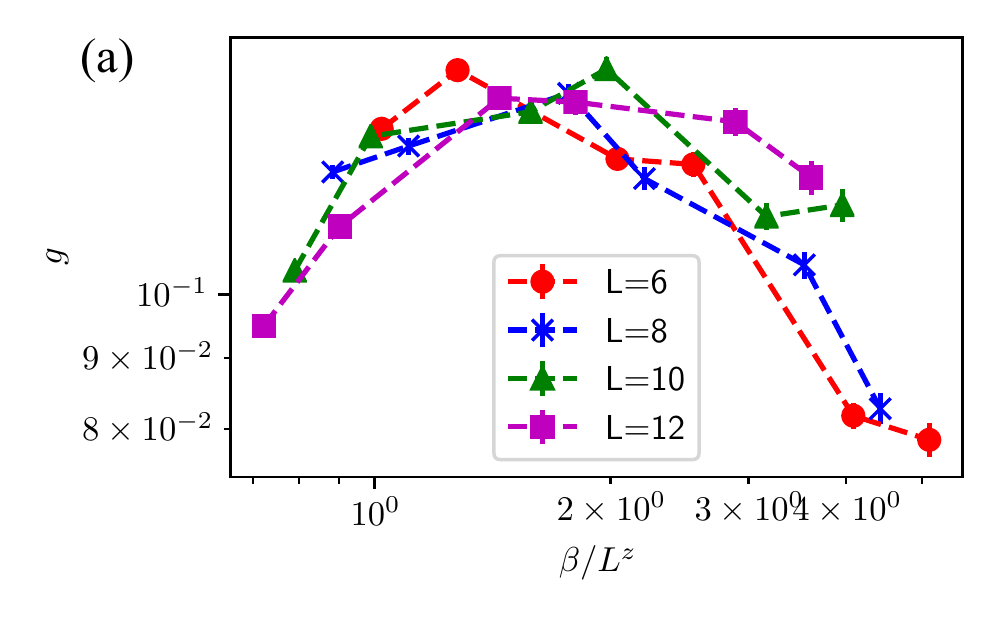}
    \includegraphics[width=.8\linewidth]{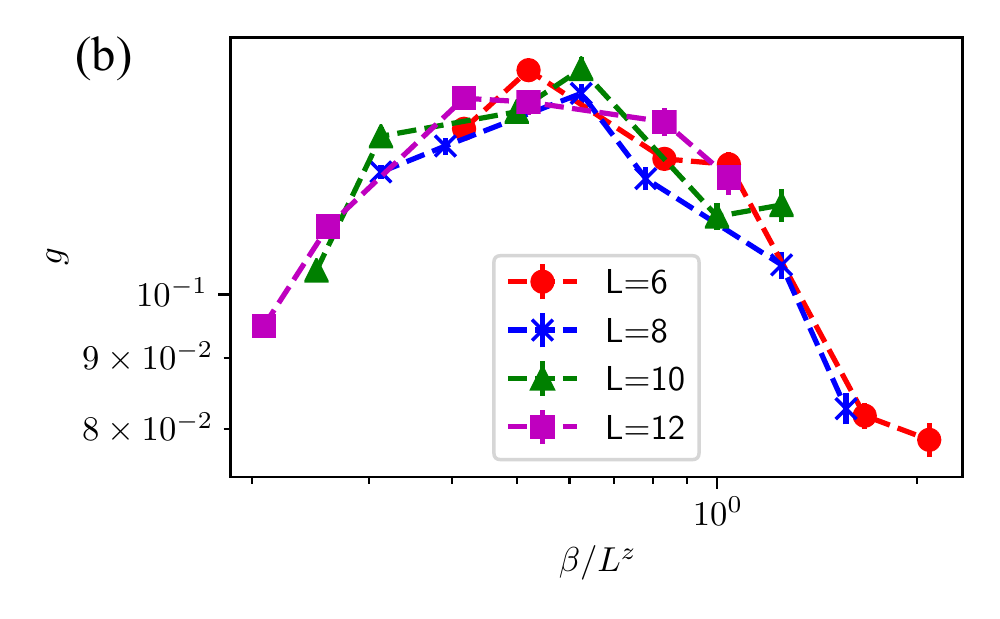}
    \includegraphics[width=.8\linewidth]{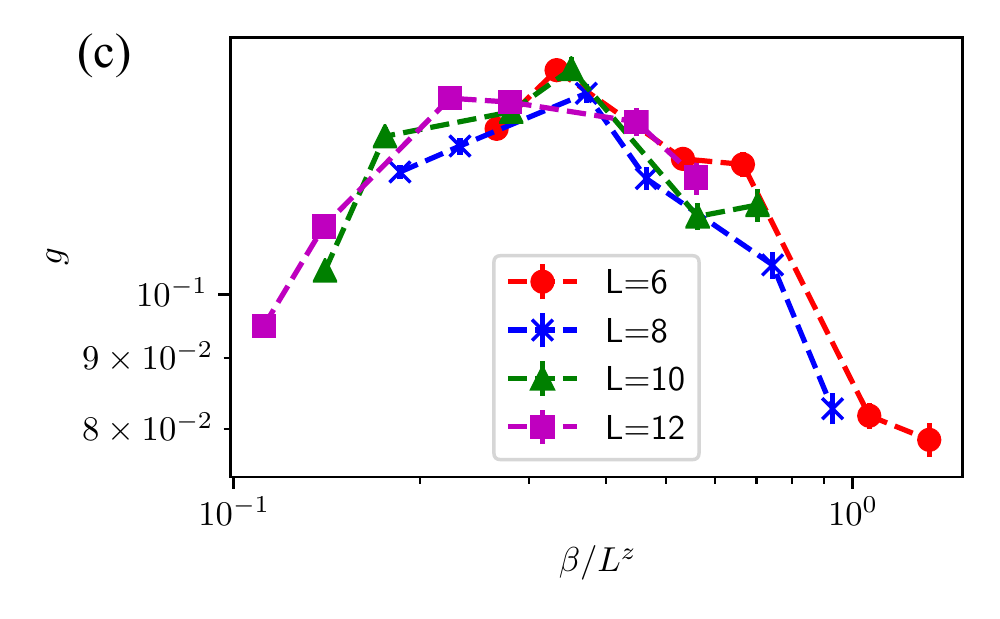}
    \includegraphics[width=.8\linewidth]{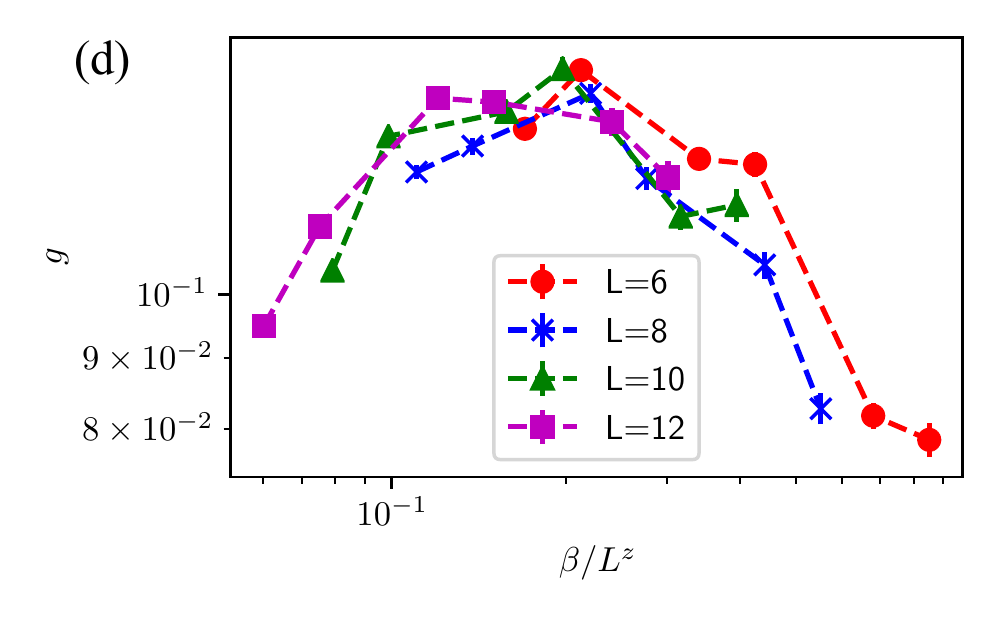}
    \caption{ Binder ratio $g$ with different sizes as a function of $\beta/L^z$ at the critical point. The dynamical critical exponent $z$ is set to $z=0.5$ (a), $z=1.0$ (b), $z=1.25$ (c), and $z=1.5$ (d). The plots use logarithmic scales on both horizontal and vertical axes.}
	\label{fig:qmc:z}
\end{figure}

\begin{figure}[tbp]
	\centering
    \includegraphics[width=.9\linewidth]{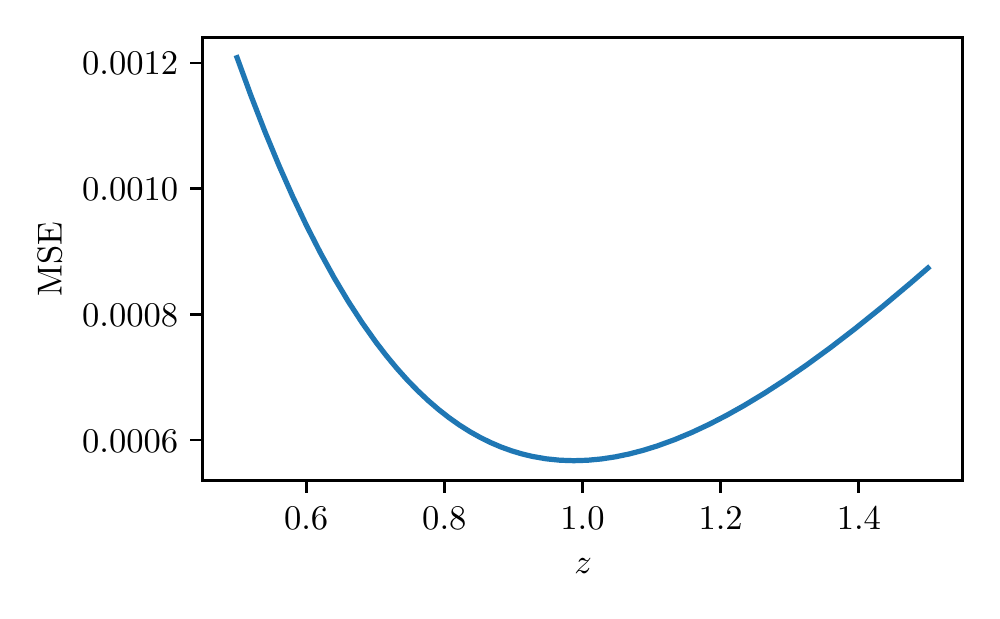}
    \caption{Mean square error (MSE) as a function of the dynamical exponent $z$. This figure shows that MSE is minimum at around $z \simeq 1$.}
	\label{fig:qmc:zfit}
\end{figure}

We next try to estimate the dynamical exponent $z$. The finite-size scaling of the Binder ratio is given by
\begin{align}
    g \simeq \tilde{f}(L^{1/\nu}(\Gamma-\Gamma_{\rm c}), \beta/L^z).
\end{align}
Figure \ref{fig:qmc:z} shows this plot.
From these figures it is apparently difficult to determine $z$ without large ambiguities, and therefore we instead employ the following ``mean square error" criteria. Figure \ref{fig:qmc:z} suggests that the function $\tilde{f}(\beta/L^z)$ can be approximated by a quadratic form, 
\begin{align}
    \ln g = a \left(\ln  \left(\frac{\beta}{L^z}\right)\right)^2
    + b \ln \left(\frac{\beta}{L^z}\right) +c,
\end{align}
where $a<0$ as seen in Fig.~\ref{fig:qmc:z}. By fitting the parameters $a ,b$ and $c$ to minimize the mean square error between the quadratic function and the actual data, we estimate the optimal parameters $a ,b$ and $c$ and the corresponding mean square error. The mean square error is depicted in Fig.~\ref{fig:qmc:zfit}, which shows $z \simeq 1$ is the best estimate, consistently with a previous study \cite{Ikegami1998}.

\section{Calibration of individual flux bias}
\label{sec:appendix_calibration}
We describe the method of calibration of qubits in the D-Wave device and its consequences (see also Ref.~\cite{King2018}). Without any calibrations, as shown in Fig.~\ref{fig:dwavenoise:calib}, some of the flux qubits tend to have the spin-up direction while others tend to have the spin-down direction even without interactions and longitudinal field. This intrinsic bias can be modeled as an effective longitudinal local field, and we try to  eliminate its effect by calibration. On the D-Wave device, this effective local field can be canceled to some extent through the ``flux-biases" option, and the task is to search the optimal flux bias for each flux qubit such that the average of a single isolated spin becomes zero.  The binary search algorithm can be used to find the optimal flux bias for each qubit. An example is listed in Algorithm \ref{alg:dwavenoise:BS}.
\begin{algorithm}[H]
    \caption{Binary search algorithm for finding the optimal flux bias for individual qubit}
    \label{alg:dwavenoise:BS}
    \begin{algorithmic}[1]
        \State $h_{i}^{\mathrm{flux}} \gets$ flux bias of $i$th flux qubit
        \State $h_{i}^{\mathrm{up}} \gets$ initial guess (positive value) of upper bound of flux bias of $i$th flux qubit
        \State $h_{i}^{\mathrm{low}} \gets$ initial guess (negative value) of lower bound of flux bias of $i$th flux qubit
        \State $N \gets$ system size
        \State $N_{\mathrm{rep}} \gets$ number of repetitions
        \Statex
        \Function{dwave}{$\{h_{i}^{\mathrm{flux}}\}$}
            \State set flux bias of $i$th qubit to $h_{i}^{\mathrm{flux}}$ for each qubit
            \State set interactions $J_{ij}$ and local fields $h_i$ to zero
            \State perform anneal-pause-quench protocol on the D-Wave machine
            \State calculate thermal average of spin $m_i$ for each qubit 
            \State \Return $\{m_i\}$
        \EndFunction{}
        \Statex
        \Function{binary-search}{$\{h_{i}^{\mathrm{up}}\}, \{h_{i}^{\mathrm{down}}\}$}
            \Loop{}
                \State $\{m_i\} = $ \Call{dwave}{$\{h_{i}^{\mathrm{up}}\}$}
                \If{$\frac{1}{N}\sum_{i=1}^{N}m_i > 0.5$}
                    \State \textbf{break}
                \EndIf
                \For{$i = 1$ to $N$}
                    \State $h_{i}^{\mathrm{up}} \gets 2h_{i}^{\mathrm{up}}$ 
                \EndFor{}
            \EndLoop
            \Statex
            \Loop{}
                \State $\{m_i\} = $ \Call{dwave}{$\{h_{i}^{\mathrm{down}}\}$}
                \If{$\frac{1}{N}\sum_{i=1}^{N}m_i < -0.5$}
                    \State \textbf{break}
                \EndIf
                \For{$i = 1$ to $N$}
                    \State $h_{i}^{\mathrm{down}} \gets 2h_{i}^{\mathrm{down}}$ 
                \EndFor{}
            \EndLoop
            \Statex
            \For{$r = 1$ to $N_{\mathrm{rep}}$}
                \For{$i = 1$ to $N$}
                    \State $(pivot)_i \gets (1/2)(h_{i}^{\mathrm{up}} + h_{i}^{\mathrm{down}})$
                \EndFor{}
                \State $\{m_i\} = $ \Call{dwave}{$\{(pivot)_i\}$}
                \For{$i = 1$ to $N$}
                    \If{$m_i > 0$}
                        \State $h_{i}^{\mathrm{up}} \gets (pivot)_i$
                    \Else
                        \State $h_{i}^{\mathrm{down}} \gets (pivot)_i$
                    \EndIf
                \EndFor{}
            \EndFor{}
            \Statex
            \State $ \{h_{i}^{\mathrm{flux}}\} = \{(1/2)(h_{i}^{\mathrm{up}}+h_{i}^{\mathrm{down}})\}$
            \State \Return{$\{h_{i}^{\mathrm{flux}}\}$}
        \EndFunction{}
    \end{algorithmic}
\end{algorithm}

\subsection*{Zero interactions}
\begin{figure}[tbp]
	\centering
	\includegraphics[width=.9\linewidth]{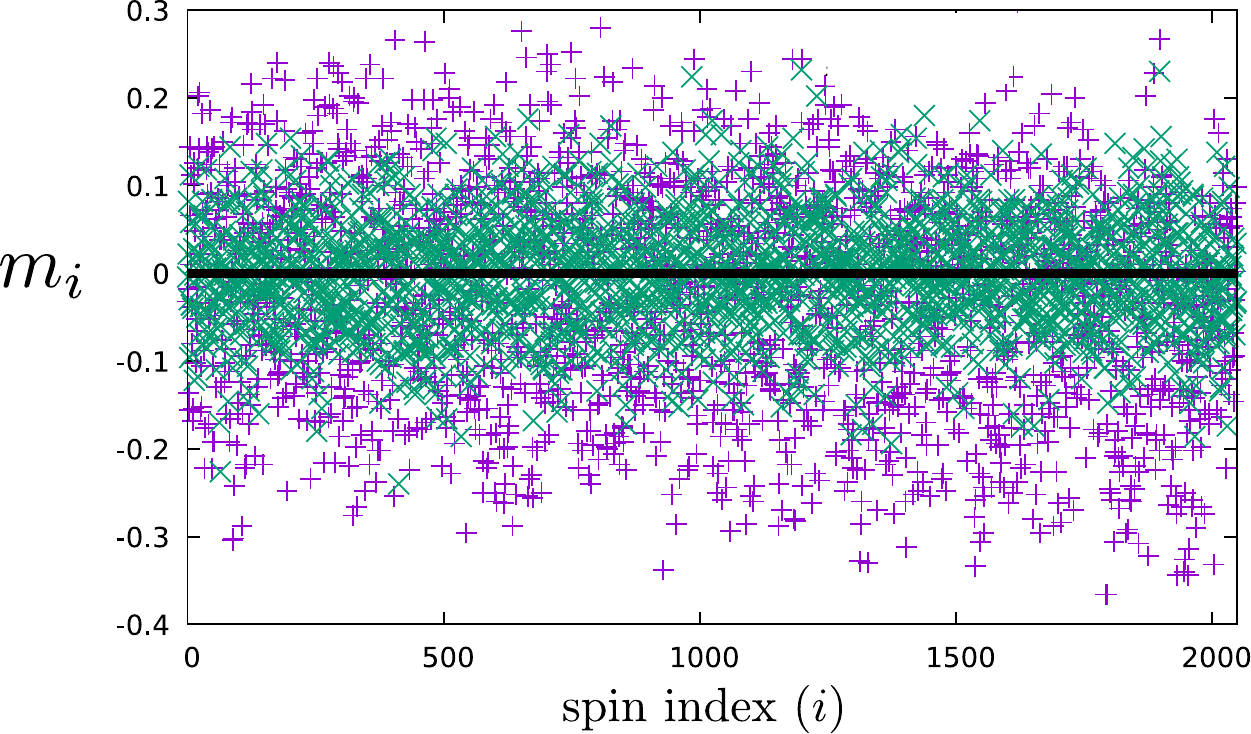}
	\caption{ Average spin configuration for each site over 100 runs of annealing. Purple dots (denoted by ``$+$'') show the result without applying error mitigation and green dots (denoted by ``$\small{\times}$'') shows the one with error mitigation.}
	\label{fig:dwavenoise:calib}
\end{figure}

We next show how these error mitigation techniques affect the data from the D-Wave device.  In the case of no interactions between qubits, We first prepare the Hamiltonian with all interactions $J_{ij}$ set to zero. We next measure the spin configuration for each qubit averaged over 100 runs of annealing, 
\begin{align}
    m_i = \sum_{a=1}^{100}\sigma_i^{a},
\end{align}
which is expected to take the value near zero.
Figure \ref{fig:dwavenoise:calib} shows $m_i$ for each site before and after applying the error mitigation techniques.
Purple dots and green dots show the result before and after applying calibration, respectively. We observe that $m_i$ of the data without error calibration tend to have values far from zero even if the interactions are set to zero. After error calibration, $m_i$ has much smaller deviations and the values are closer to zero than the case without calibration.

\subsection*{Diluted Chimera graph}

\begin{figure*}[tbp]
	\centering
     \includegraphics[width=.4\linewidth]{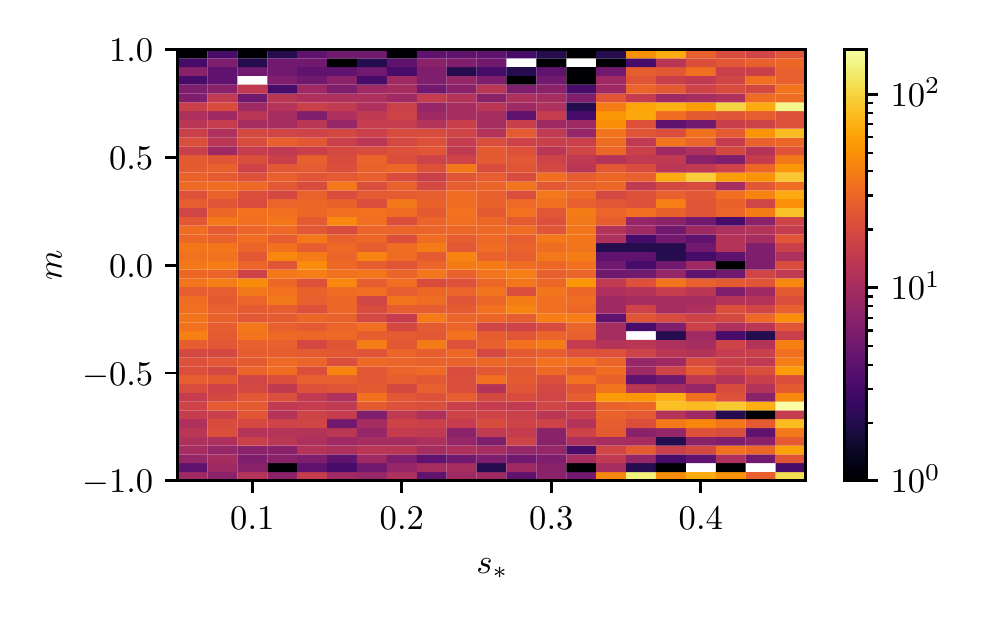}
    \includegraphics[width=.4\linewidth]{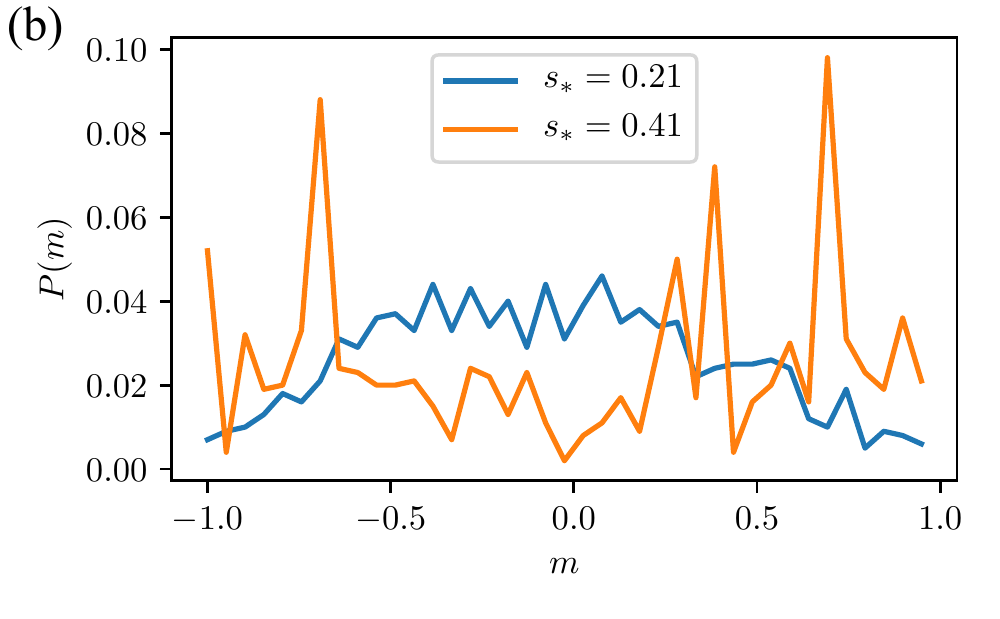}
    \caption{ (a) Histogram (heat map) of magnetization without calibration as a function of the pause point.  The notation is the same as in Fig.~\ref{fig:dwave:mhist}. (b) Histogram of the magnetization without calibration with the pause point $s_{*}$ is fixed to 0.21 (paramagnetic phase) and 0.41 (ferromagnetic phase).}
	\label{fig:dwavenoise:mhist}
\end{figure*}

Next we apply the above technique to the diluted Chimera graph and see how the result changes by calibration. Figure \ref{fig:dwavenoise:mhist} shows the histogram of the magnetization without calibration.
We see that there are unphysical multiple peaks in the ferromagnetic region, where the magnetization is expected to have peaks close to $\pm 1$. Although we  see a symptom of phase transition around $s_{*} = 0.35$, it is quite hard to reliably extract information from this set of data.

\begin{figure*}[tbp]
	\centering
    \includegraphics[width=.4\linewidth]{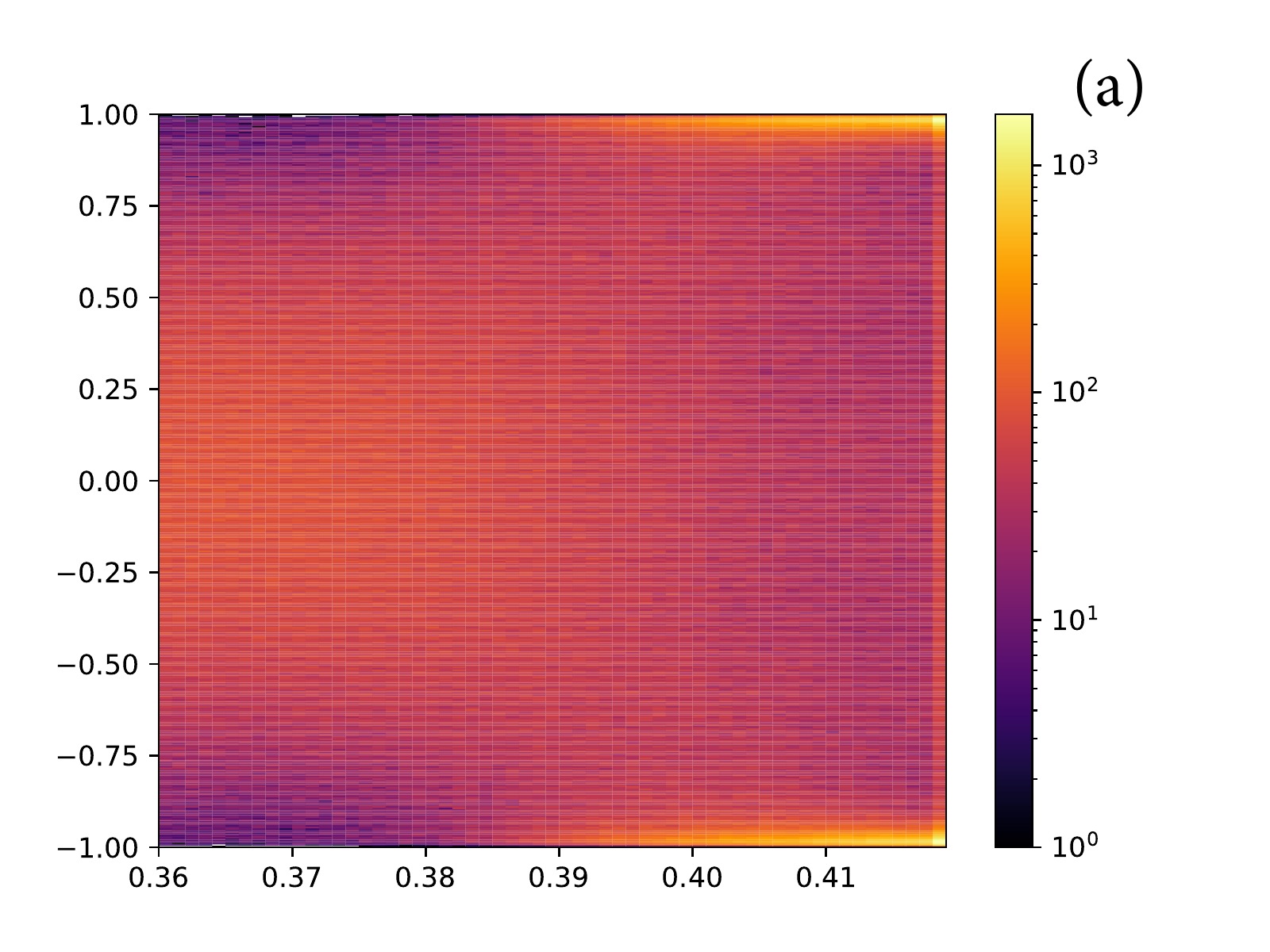}
    \includegraphics[width=.4\linewidth]{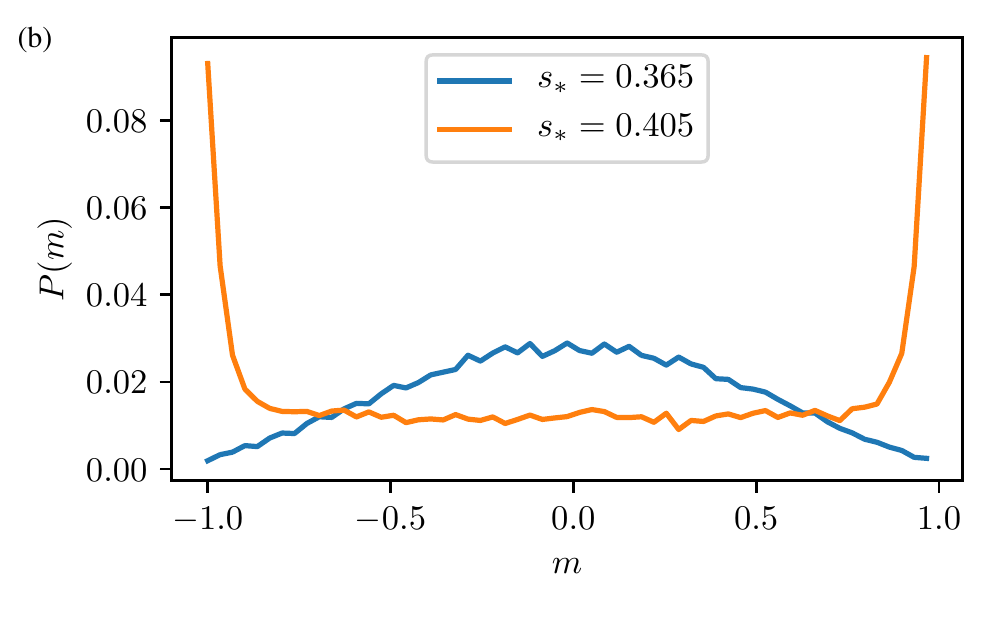}
    \caption{ (a) Histogram (heat map) of magnetization after calibration. (b) Histogram of the magnetization after calibration with the pause point $s_{*}$ fixed to 0.21 (paramagnetic phase) and 0.41 (ferromagnetic phase). These figures are the same as Fig.~\ref{fig:dwave:mhist}.}
	\label{fig:dwavenoise:mhistafter}
\end{figure*}

Figure \ref{fig:dwavenoise:mhistafter} shows the data after calibration.
We clearly observe peaks only at $\pm 1$ in the ferromagnetic region, as they should.  From these data we confirm that the calibration process is effective and indispensable for reliable quantum simulations on the D-Wave device. 

\bibliographystyle{apsrevtitle}
\bibliography{main.bib}
\end{document}